\documentclass[graybox]{svmult}

\usepackage{type1cm}        
\usepackage{makeidx}         
\usepackage{graphicx}        
\usepackage{multicol}        
\usepackage[bottom]{footmisc}

\usepackage{newtxtext}       %
\usepackage{newtxmath}       

\makeindex             

\newcommand{\om}{\Omega_{\rm m}}
\newcommand{\ok}{\Omega_{\rm k}}
\newcommand{\ol}{\Omega_\Lambda}
\newcommand{\oc}{\Omega_{\rm c}}
\newcommand{\ob}{\Omega_{\rm b}}
\newcommand{\as}{A_{\rm s}}
\newcommand{\ns}{n_{\rm s}}

\begin{document}

\title*{The establishment of the Standard Cosmological Model through observations}
\author{Ricardo  Tanaus\'u G\'enova-Santos}
\institute{Instituto de Astrof\'{\i}sica de Canarias, E-38200 La Laguna, Tenerife, Spain \and Departamento de Astrof\'{\i}sica, Universidad de La Laguna (ULL), E-38206 La Laguna, Tenerife, Spain, \email{rgs@iac.es}
}
%
%
\maketitle

\abstract{Over the last decades, observations with increasing quality have revolutionized our understanding of the general properties of the Universe. Questions posed for millenia by mankind about the origin, evolution and structure of the cosmos have found an answer. This has been possible mainly thanks to observations of the Cosmic Microwave Background, of the large-scale distribution of matter structure in the local Universe, and of type Ia supernovae that have revealed the accelerated expansion of the Universe. All these observations have successfully converged into the so-called "concordance model". In spite of all these observational successes, there are still some important open problems, the most obvious of which are what generated the initial matter inhomogeneities that led to the structure observable in today's Universe, and what is the nature of dark matter, and of the dark energy that drives the accelerated expansion.
In this chapter I will expand on the previous aspects. I will present a general description of the Standard Cosmological Model of the Universe, with special emphasis on the most recent observations that have allowed us to consolidate this model. I will also discuss the shortfalls of this model, its most pressing open questions, and will briefly describe the observational programmes that are being planned to tackle these issues.\newline\indent}

\section{Cosmological models}
\label{sec:cosmological_models}

\subsection{First cosmological views and models}
\label{sec:first_models}
Since its beginnings, mankind has questioned about the origin, history and evolution of the Cosmos. During the 5$^{\rm th}$ century B.C. classical philosophers and thinkers like Anaxagoras, Leucippus of Mileto or Democritus hypothesised about the main constituents of the Universe. After that, the first models attempting to describe the closest observable Universe (the orbits of the Solar System planets) appeared. Aristotle (4$^{\rm th}$ century B.C.) proposed the geocentric model and a finite and static Universe built up of fire, air, water and earth. In the 2$^{\rm th}$ century B.C. Ptolomy further developed this model by introducing the concept of epicycles. Regrettably this was the prevalent model during more than 17 centuries, and throughout the full Middle Ages. In 1543 Nicolaus Copernicus caused a big breakthrough with the publication of his book {\it De revolutionibus orbium coelestium}, where he suggested that the Sun, rather than the Earth, was the centre of the Solar System. This idea had initially been put forward by Aristarchus of Samos in the 3$^{\rm rd}$ century B.C., although it had been totally ignored. Copernicus theory marked a major event in the history of science, triggered the Copernican Revolution, and made a pioneering contribution to the so-called Scientfic Revolution, which together with other events and developments in other fields in turn marked the emergence of the modern science during the early modern period. Later in 1584 Giordano Bruno went further and dared to suggest that not even the Solar System was the centre of the Universe, with the Sun being just one more star amongst many others. This had well-known terribly bad consequences for him. In 1605 Johannes Kepler introduced refinements on the Copernican model, such as proposing elliptical instead of circular planet orbits, based on observations carried out jointly with Tycho Brahe. We can probably say that these developments marked the transition of Cosmology from a philosophical and speculative to a more scientific discipline, which involved an attempt to perform the first mathematical description of the Cosmos based on the observations.

It is often considered that the Scientific Revolution that started with the ideas of Copernicus was culminated with the publication by Isaac Newton of his ``Mathematical Principles of Natural Philosophy'' in 1687, where he formulated the laws of motion and universal gravitation, thence lying the foundation of classical mechanics. Newton proposed a finite and static Universe, in which matter is uniformly distributed. He thought that gravitational forces between planets should lead to an unstable Solar System and, sticking to his religious beliefs, he stated that God's intervention made it an stable system. This idea was challenged 100 years later by Pierre-Simon Laplace, who thought that mathematics should be able to explain the motion of planets without the need of God. In the same way as Newton did, Albert Einstein naturally attempted to apply his theory to cosmology, this time the theory of General Relativity, published in 1915. Einstein also embraced the prevailing idea at that time of a dynamically static Universe. Instead of resorting to God, in this case he modified his original field equations by the introduction of the cosmological constant, $\Lambda$, which he viewed as a repulsive form of gravity that kept the Universe stable. This changed dramatically with the discovery of the Universe expansion by Edwin Hubble in 1929, which led Einstein to recognise as his biggest blunder the assumption of a static universe and the introduction of the cosmological constant.

\subsection{Big Bang Cosmology}
\label{sec:big_bang}

Although confirmed observationally in 1929 by Edwin Hubble, the idea of an expanding universe had been proposed in 1922 by Alexander Friedmann, based on Einstein's fundamental equations. In 1927 Georges Lema\^{\i}tre, backed by the first observations of Edwin Hubble, concluded that the Universe was in fact expanding, and this naturally drove him to propose for the first time the idea of a Big Bang (he initially coined his idea ``primeval atom''). Later in 1946 George Gamow \cite{gamow_46} proposed a hot Big Bang model to explain the primeval build-up of elements heavier than hydrogen. He realised that reaching the binding energies of those nuclei required temperatures of the order of $\sim 10^9-10^{10}$~K. According to this model the Universe was sufficiently hot and dense in its beginnings, and later expanded out and cooled down, in such a way that now it should have an average non-zero temperature. This remnant temperature should show up as a background radiation with a temperature of $\sim 5$~K, the so-called ``Cosmic Microwave Background'' (CMB), according to the prediction by Ralph A. Alpher and Robert Herman \cite{alpher_48}. 

At that time, a competing alternative was the ``Steady State Model'', which was proposed by three prominent physicists: Hermann Bondi, Thomas Gold and Fred Hoyle. This model proposed that the observed expansion of the Universe was associated with a spontaneous and continuous creation of matter, which led to a Universe with a fixed average density \cite{bondi_48}. The existence of the Cosmic Microwave Background could not fit by any means in this theory, and therefore its discovery in 1964 by Arno Penzias and Robert Wilson\footnote{Note that, even before this, the CMB had been indirectly detected by \cite{adams_41} through the local excitation of CN molecules in our Galaxy. This excitation was attributed to some kind of ``unknown'' radiation with temperature $\sim 2.3$~K.} \cite{penzias_65} led to the refutal of this model, and to the final recognition of the Big Bang Model. Ever since this has been the prevailing model to explain the beginning and later expansion of our Universe. The three main, and independent, pillars supporting nowadays this theory are: i) the expansion of the Universe; ii) the abundance of light elements; and iii) the CMB.

\subsection{The Standard Model of Cosmology}
\label{sec:smc}

All the previous developments and discoveries led to the establishment of the what we nowadays call the ``Standard Model of Cosmology'' (SMC). This model is supported mainly on General Relativy and by the validity of the ``Cosmological Principle'', which states that the Universe is homogeneous and isotropic on large scales ($\gtrsim 100$~Mpc). The geometry of a homogeneous and isotropic Universe in a 4-dimensional (space and time) system is described by the Robertson-Walker metric,
\begin{equation}
ds^2 = c^2 dt^2 - a^2(t) \left[\frac{dr^2}{1-kr^2}+r^2\left(d\theta^2+{\rm sin}^2\theta d\phi^2\right) \right] ~~,
\label{eq:rw_metric}
\end{equation}
where $a(t)$ is the scale factor and $k=-1, 0, +1$ respectively for an open, flat or closed Universe. Alexander Friedmann used this metric to solve the Einstein's field equations of General Relativity, leading to the so-called Friedmann equations:
\begin{equation}
\left(\frac{\dot{a}}{a}\right)^2 = \frac{8\pi G}{3}\rho-\frac{k c^2}{a^2}+\frac{\Lambda c^2}{3} ~~,
\label{eq:friedmann_model_1}
\end{equation}
\begin{equation}
\frac{\ddot{a}}{a} = -\frac{4\pi G}{3}\left( \rho+\frac{3p}{c^2}\right)+\frac{\Lambda}{3} c^2 ~~,
\label{eq:friedmann_model_2}
\end{equation}
where $p$ and $\rho$ are respectively pressure and density, $G$ is the gravitational constant, $c$ the speed of light and $\Lambda$ is the ``cosmological constant''. Equation~\ref{eq:friedmann_model_1} gives the expansion rate, whereas equation~\ref{eq:friedmann_model_2} gives the acceleration of the expansion. By defining the following dimensionless mass, curvature and dark-energy (or vacuum energy) density parameters:
\begin{equation}
\om=\frac{8\pi G}{3 H_0^2}\rho_0 ~~~~,~~~~ \ok=-\frac{k c^2}{a_0^2 H_0^2}~~~~{\rm y}~~~~
\ol = \frac{\Lambda c^2}{3 H_0^2} ~~~~,
\label{eq:adimen_param}
\end{equation}
where $H_0=(\dot{a_0}/a_0)$ denotes the present value of the Hubble parameter, the first Friedmann equation (\ref{eq:friedmann_model_1}) can be written at $t_0$ (current time) as\footnote{A more detailed explanation of the derivation of the Friedmann equations can be found in classical references like \cite{weinberg_72}.}
\begin{equation}
\om+\ok+\ol=1 ~~.
\label{eq:friedmann_model_1b}
\end{equation}
The total density of the Universe can be written as $\Omega_{\rm tot}=\Omega_{\rm m}+\Omega_\Lambda=1-\Omega_{\rm k}$, and the critical density is the value required for a flat geometry ($\Omega_{\rm k}=0\Rightarrow \Omega_{\rm tot}=1$), $\rho_c=3H_0^2/(8\pi G)=1.879~h^2\times 10^{-29}$~g~cm$^{-3}$, in such a way that $\Omega_{\rm m}=\rho_0/\rho_c$. This flat geometry is confirmed by observations, as we will see in section~\ref{sec:obs_probes}.

We now have evidence for a non-zero dark energy component (see section~\ref{sec:obs_probes}), with $\Omega_\Lambda\approx 0.69$. The other dark component of the Universe is coined ``dark matter'', the first evidence of which date from 1933 when Fritz Zwicky \cite{zwicky_33} encountered a discrepancy on the the total mass of the Coma cluster inferred through the velocity dispersion of individual member galaxies and through its emitted light. Another well-known evidence came later through the analysis of the rotation curves of galaxies \cite{rubin_80}, while nowadays we have even further compelling pieces of evidence of this component derived from gravitational lensing, the large-scale structure of the Universe, or the CMB, as we will see in section~\ref{sec:obs_probes}. These observations also agree on that this component should be ``cold''. Therefore, the matter density is decomposed into a cold-dark-matter component and a baryonic component, $\om=\ob+\oc$, and we know that $\oc\approx 0.27$ (see section~\ref{sec:obs_probes}). Therefore $\ol+\oc\approx 0.95$, and then we can say that only $5\%$ of the total matter-energy budget of the Universe is ordinary matter, the rest being dark matter and dark energy. For this reason the current SMC has been coined $\Lambda$CDM.

The first Friedmann equation (\ref{eq:friedmann_model_1b}) describes the composition and the geometry of the Universe. Based on compelling observational pieces of evidence, in the $\Lambda$CDM model the universe is considered to be flat, so $\ok=0$ is fixed. Then, given the existence of the two matter components, we end up with two independent parameters, which are usually considered to be $\ob$ and $\om$, while the density of dark energy is derived from equation~\ref{eq:friedmann_model_1b} as $\ol = 1-\ob-\oc$. A third important free parameter in the model, which describes the evolution of the Universe, is the Hubble constant, $H_0$. Nowadays it is common practice to use as a free parameter the angular acoustic scale at recombination, $\theta_{\rm MC}$, instead of $H_0$, because it has less correlations with other cosmological parameters.

The previous three parameters accounting for the composition and evolution of the Universe must be complemented with three additional parameters. Two of them describe the initial conditions of the primordial density perturbations, whose later evolution leads to structure formation. More specifically these parameters are the amplitude $\as$ and the spectral index $\ns$ of the primordial power spectrum of scalar perturbations,
\begin{equation}
P(k) = A_{\rm s}\left(\frac{k}{k_0}\right)^{n_{\rm s}-1}~~,
\label{eq:spectrum_scalar}
\end{equation}
where $k_0$ is an arbitrary pivot scale.

Astrophysical parameters, related to different processes like radiation transfer, ionisation or recombination, must also be considered. The most important of these, which becomes the sixth parameter in the $\Lambda$CDM model, is the optical depth of Thomson scattering to recombination, $\tau$, which is defined as
\begin{equation}
\tau = \sigma_{\rm T}\int_0^{z_{\rm rec}} n_{\rm e}(z')~dz'~~,
\label{eq:tau}
\end{equation}
where $\sigma_{\rm T}$ is the Thomson cross section and $n_{\rm e}(z)$ the electron density at redshift $z$.

We commented earlier that the geometry parameter, $\ok$, is usually considered as a fixed parameter in this parametrisation. Another important parameter that is usually fixed is the CMB temperature (or equivalently the radiation density), thanks to the outstanding precision with which this parameter can be measured by fitting the CMB frequency spectrum, $T_0=2.7260\pm 0.0013$~K \cite{fixsen_09}.

To sum up, we have a 6-parameter model, $\{\ob,\oc,\theta_{\rm MC},A_{\rm s},n_{\rm s},\tau\}$, which describes the global properties of our Universe, and which provides an astonishingly good description of the available data, as we will see in section~\ref{sec:obs_probes}.

\subsection{Extensions to the $\Lambda$CDM model}
\label{sec:extensions}
As we have seen before in section~\ref{sec:smc}, the $\Lambda$CDM model has 6 independent parameters, and provides an excellent fit to all kind of available data. Two other parameters, curvature $\ok$ and the radiation density $T_0$, are very well constrained by observations and are left fixed. Different extensions to this model, involving new parameters, can be considered. We will now briefly describe some of these extensions.

\subsubsection{Early-Universe physics and initial conditions}
A complete cosmological model must include an statistical description of the tiny initial perturbations that gave rise to the structure we see in today's Universe. The physical framework that provides this is inflation \cite{starobinsky_79,guth_81}, according to which the early Universe underwent a brief period of accelerated expansion during which quantum fluctuations were stretched out to become the classical fluctuations that we see today. Inflationary cosmology provides an elegant explanation of the key features of our Universe: flat geometry ($\ok=0$), adiabatic, nearly scale-invariant ($n_{\rm s}\lesssim 1$, very close to unity), and Gaussian perturbations on all scales. However, some other phenomenological models of inflation (see \cite{linde_08} for a review), or even alternatives to inflation, predict departures from these conditions, or distinctive signatures that may be seen in the data. 

{\bf Scale-dependent primordial fluctuations.}
In the base $\Lambda$CDM model primordial fluctuations are modelled by a pure power-law with spectral index $n_{\rm s}$ (equation~\ref{eq:spectrum_scalar}). The simplest inflationary models predict a nearly scale-invariant spectrum, $n_{\rm s}\approx 1$. However, there are inflationary models that allow for a running of the spectral index, ${\rm d}n_{\rm s}/{\rm d ln}k$:
\begin{equation}
P(k) = A_{\rm s}\left(\frac{k}{k_0}\right)^{n_{\rm s}-1+(1/2)({\rm d}n_{\rm s}/{\rm d ln}k){\rm ln}(k/k_0)}~~.
\label{eq:spectrum_scalar_running}
\end{equation}

\begin{figure}
\sidecaption
\includegraphics[scale=.31]{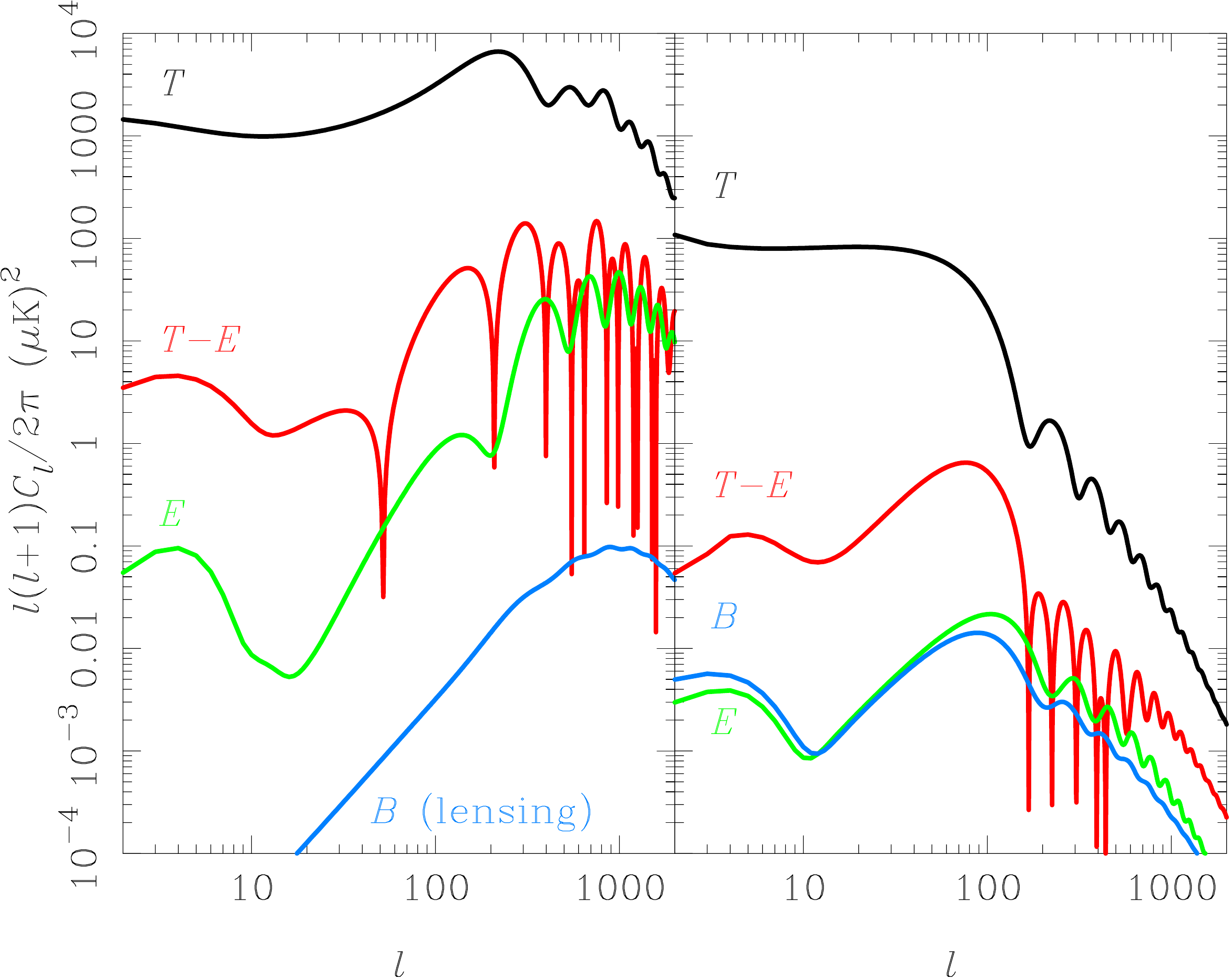}
\caption{Temperature (black), E-mode (green), B-mode (blue) and TE cross-correlation (red) CMB power spectra produced by scalar perturbations (left) and tensor perturbations (gravitational waves; right). The amplitude of the tensor perturbations has been fixed at $r=0.22$. In the left-hand panel the B-mode spectrum induced by weak gravitational lensing is also depicted. Figure taken from \cite{challinor_13}.}
\label{fig:scalar_tensor_ps}
\end{figure}

{\bf Tensor perturbations.}
Although in the base $\Lambda$CDM model perturbations are assumed to be purely scalar modes, primordial tensor perturbations (gravitational waves) should also be originated during inflation. The tensor-mode spectrum is usually described as
\begin{equation}
P_{\rm t}(k) = A_{\rm t}\left(\frac{k}{k_0}\right)^{n_{\rm t}}~~.
\label{eq:spectrum_tensor}
\end{equation}
The ratio between the amplitude of the tensor to the scalar modes, at the reference scale $k_0$, is denoted as $r=A_{\rm t}/A_{\rm s}$. This parameter is proportional to the fourth power of the energy scale of inflation. Slow-roll inflation models satisfy the consistency relation $n_{\rm t}=-r/8$. As it is shown in Figure~\ref{fig:scalar_tensor_ps}, tensor modes contribute to temperature anisotropies on the large scales (low multipoles, $\ell$), but are completely overshadowed by the scalar modes. On the other hand, a confirmation of the existence of tensor modes could be unambiguously achieved by detecting the so-called B-mode anisotropy at large angular scales (see section~\ref{sec:cmb}).

{\bf Spatial curvature.}
The base $\Lambda$CDM model assumes Friedmann-Lema\^{\i}tre-Robertson-Walker metric with a flat geometry. As we will see in section~\ref{sec:obs_probes} all current observations are consistent with $\ok=0$ to very high accuracy. In fact, providing an explanation for the flatness of our observed Universe was one of the main motivations of the inflationary cosmology. However, there are specific inflationary models that generate open \cite{linde_99} or closed \cite{linde_03} universes. The detection of a deviation from $\ok=0$ would have profound implications both for inflationary models and for fundamental physics.

{\bf Isocurvature perturbations.}
As it was said above, in $\Lambda$CDM the primordial perturbations are considered to be adiabatic. This is what observations currently show, and in fact it is a key prediction of single-field inflation. On the other hand, isocurvature fluctuations (fluctuations in the relative energy densities of different species at a fixed total density) could be produced by sub-dominant fields during the inflation era. Detecting them could then yield important insights on physics at the energy scale of inflation.

\subsubsection{Dark energy}
\label{sec:dark_energy}
The late-time accelerated expansion of the Universe is still considered one of the most mysterious aspects of standard cosmology. In the base $\Lambda$CDM model the acceleration is provided by a cosmological constant $\Lambda$ that is introduced in Einstein equations of General Relativity (see section~\ref{sec:smc}). However there are many other alternatives, like a dynamical dark energy based in scalar fields or modifications of general relativity on cosmological scales (see \cite{tsujikawa_10}). While a cosmological constant has equation of state $w=p/\rho=-1$, scalar field models usually have a time varying $w$ with $w\ge -1$, while other models have $w<-1$. It is then important to try to constrain $w$ to get insight on the nature of dark energy.

\subsubsection{Neutrinos}
In $\Lambda$CDM neutrinos are assumed to be massless. Nonetheless the flavour oscillations observed in solar and atmospheric neutrinos indicate that neutrinos should be massive, implying striking evidence of physics beyond the standard model. Constraining the value of the neutrino mass is therefore one of the key questions in fundamental physics. In addition of assuming zero mass, $\Lambda$CDM fixes the number of relativistic species to $N_{\rm eff}=3.046$ (the reason of being slightly larger than three is that the three standard model neutrinos were not completely decoupled at electron-positron annihilation). Exploring different $N_{\rm eff}$ values would be useful to test extensions of the standard model that predict the existence of new light particles.


\subsubsection{Variation of fundamental constants} 
The $\Lambda$CDM model assumes the validity of General Relativity on cosmological scales, and of the standard model of particle physics on small scales. One possible extension to this model, which may have important implications in fundamental physics, is to consider a possible variation of dimensionless constants. These include the fine-structure constant, $\alpha$, the electron-to-proton mass ratio, $\mu$, and the gravitational constant, $\alpha_{\rm g}=Gm_{\rm p}^2/\hbar c$. While a variation of $G$ can affect Friedmann equations, and also raise the issue of consistency in the overall theory of gravity, variations of the non-gravitational constants such as $\alpha$ or $\mu$ could be produced for instance if there exists a new interaction between light and atoms mediated by a new massless scalar field \cite{uzan_03}. Another important prediction of the $\Lambda$CDM scenario is the redshift evolution of the CMB temperature (or radiation density) which, under the assumption of adiabatic expansion of the Universe and photon-number conservation, should be $T_{\rm CMB}(z)=T_0(1+z)$. However, there are non-standard scenarios like a non perfectly transparent Universe, decaying vacuum cosmologies or modified gravity models, which lead to photon mixing and/or violation of photon number conservation, and therefore to a deviation from the previous standard redshift evolution.

\section{Observational probes}
\label{sec:obs_probes}

In the previous section we have described the basics of the $\Lambda$CDM model. Of course, the establishment of this model is not only a product of human thinking, but of the analysis and interpretation of huge amount of data coming from different kind of observations. Over the last decades, technological improvements in the design and development of astronomical instrumentation have led to data with increasing quality and sensitivity, allowing us to determine cosmological parameters with unprecedented accuracy. This is why we usually say that we live in the era of ``precision cosmology''. As a result of the beautiful consensus between completely different and independent kind of observations, the current $\Lambda$CDM model has been coined the ``concordance model''. We will now describe the three most important cosmological probes.

\subsection{The Cosmic Microwave Background}
\label{sec:cmb}

The CMB currently stands as the observational probe giving the tightest constraints on cosmological parameters. Since the discovery by the COBE satellite of the temperature anisotropies \cite{smoot_92}, different experiments (from the ground, balloon-borne or space satellites) have been specifically designed to measure and characterise these anisotropies with gradually finer angular resolution and sensitivity. These anisotropies (with amplitude $\sim 10^{-5}$ with respect to the mean CMB temperature) are produced by fluctuations of the baryon-photon fluid at the last-scattering surface ($z\sim 1100$) in the gravitational potential wells produced during the  inflationary period. All the statistical information of these anisotropies can be neatly encoded in the so-called ``power spectrum'', which is defined in terms of a spherical harmonics expansion:
\begin{equation}
\frac{\Delta T}{T_0}(\theta,\phi)=\sum_{\ell=2}^{\infty}\sum_{m=-\ell}^{\ell} a_{\ell m} Y_{\ell m}(\theta,\phi) ~~,
\label{eq:dt_arm_esf}
\end{equation}
where $\Delta T$ represents the CMB temperature spatial variations with respect to their mean value $T_0$. Given that under $\Lambda$CDM temperature fluctuations are Gaussian, the $a_{\ell m}$ coefficients must be random and statistically independent variables. The value of $a_{\ell m}a^{*}_{\ell m}$ averaged over the full sky gives an estimate of the power associated with each multipole $\ell$, and for this reason we define the power spectrum of the temperature variation as the variance of this quantity,
\begin{equation}
C_{\ell} = \frac{1}{2\ell+1}\sum_{m=-\ell}^{\ell} a_{\ell m} a^{*}_{\ell m} = \langle |a_{\ell m}|^{2} \rangle ~~.
\label{eq:ep_def}
\end{equation}
It is common practice to plot the quantity $\mathcal{D}_\ell=\ell(\ell+1)C_\ell/(2\pi)$.

The peaks in the power spectrum (see~Figure~\ref{fig:planck_ps}) are associated with the spatial distribution of the CMB anisotropies, and reflect the acoustic oscillations in the baryon-fluid plasma in the early Universe, which are frozen during the matter-radiation decoupling. The relative positions and amplitudes of these peaks are tightly sensitive to a number of cosmological parameters, in particular to the total energy density and to the curvature of the Universe. This motivates the interest to achieve an accurate characterisation of the CMB power spectrum. 

The BOOMERanG experiment was the first that clearly delineated the first peak of the CMB power spectrum \cite{debernardis_00}, at $\ell\approx 200$ (angular scales of $\approx 1^\circ$ on the sky), inferring the flatness of the Universe. These results were confirmed soon after that by other balloons like MAXIMA \cite{stompor_01} or ARCHEOPS \cite{benoit_03}. In the following $\sim 5$ years, ground-based experiments like VSA \cite{dickinson_04}, CBI \cite{sievers_03} or ACBAR \cite{kuo_04} were able to measure higher-order peaks out to $\ell\approx 1500$. NASA's WMAP satellite observed the full sky with angular resolution of $15$~arcmin \cite{bennett_13}, comparable to previous ground-based experiments, resulting in a significant improvement in the measurement of the first peak, and also covering the second and third peaks out to $\ell\approx 1000$. Slightly later on time, ACT \cite{fowler_10} and SPT \cite{story_13}, a new generation of ground-based experiments, reached angular resolutions of $\sim 1$~arcmin on small sky regions ($\sim 200$~deg$^2$), allowing the power spectrum to be measured out to $\ell\sim 4000$ (see Figure~\ref{fig:planck_ps}).

\begin{figure}
\includegraphics[scale=0.45]{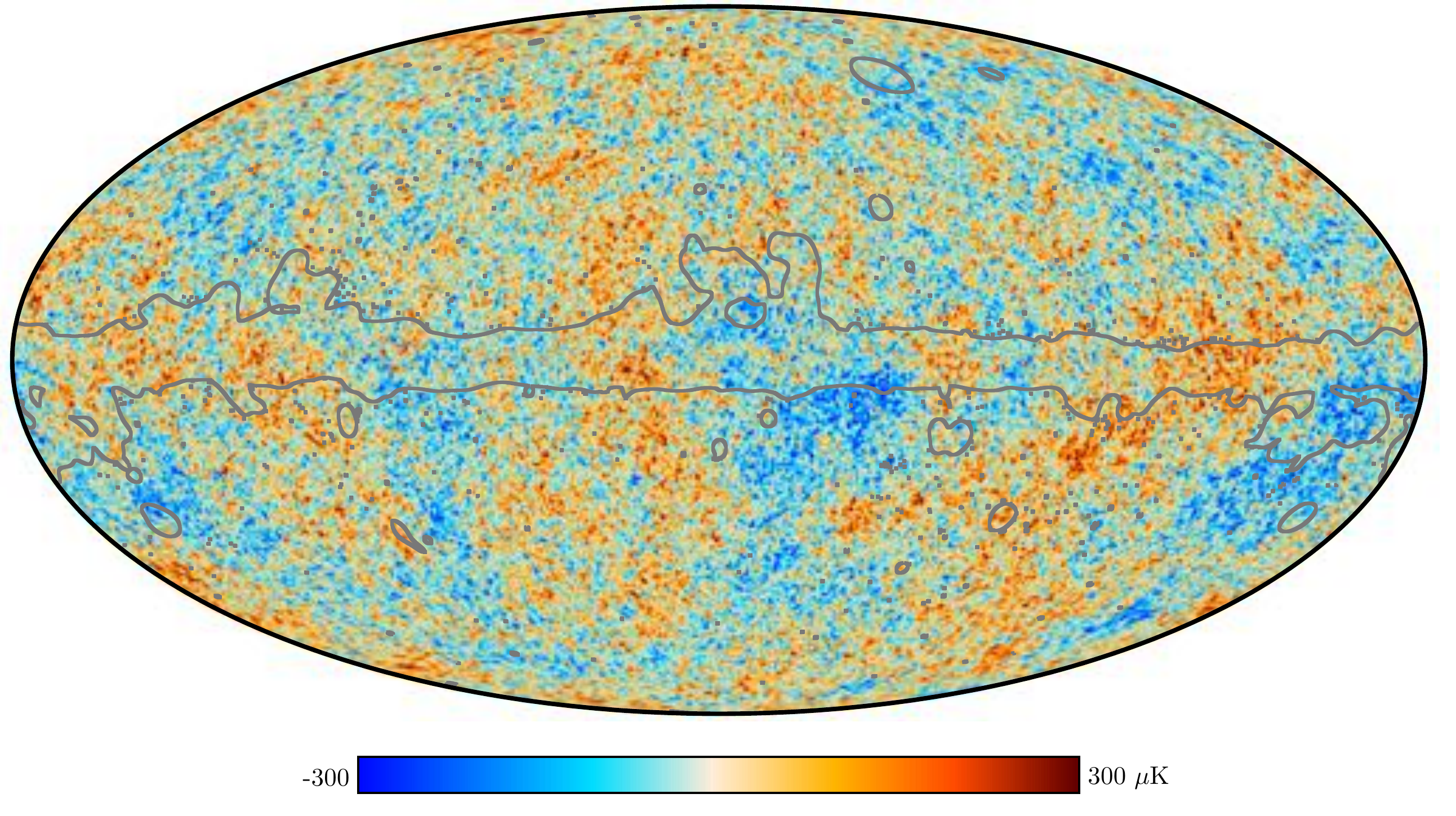}
\caption{Full-sky CMB map derived from Planck multi-frequency data \cite{planck_18_1}, projected into Galactic coordinates. The grey lines indicate the masked regions (mostly encompassing the Galactic plane) that have been excluded in the cosmological analyses. Figure taken from the Planck Legacy Archive ({\tt https://www.cosmos.esa.int/web/planck/}).
}
\label{fig:planck_cmb_map}
\end{figure}

\begin{figure}
\sidecaption
\includegraphics[scale=0.25]{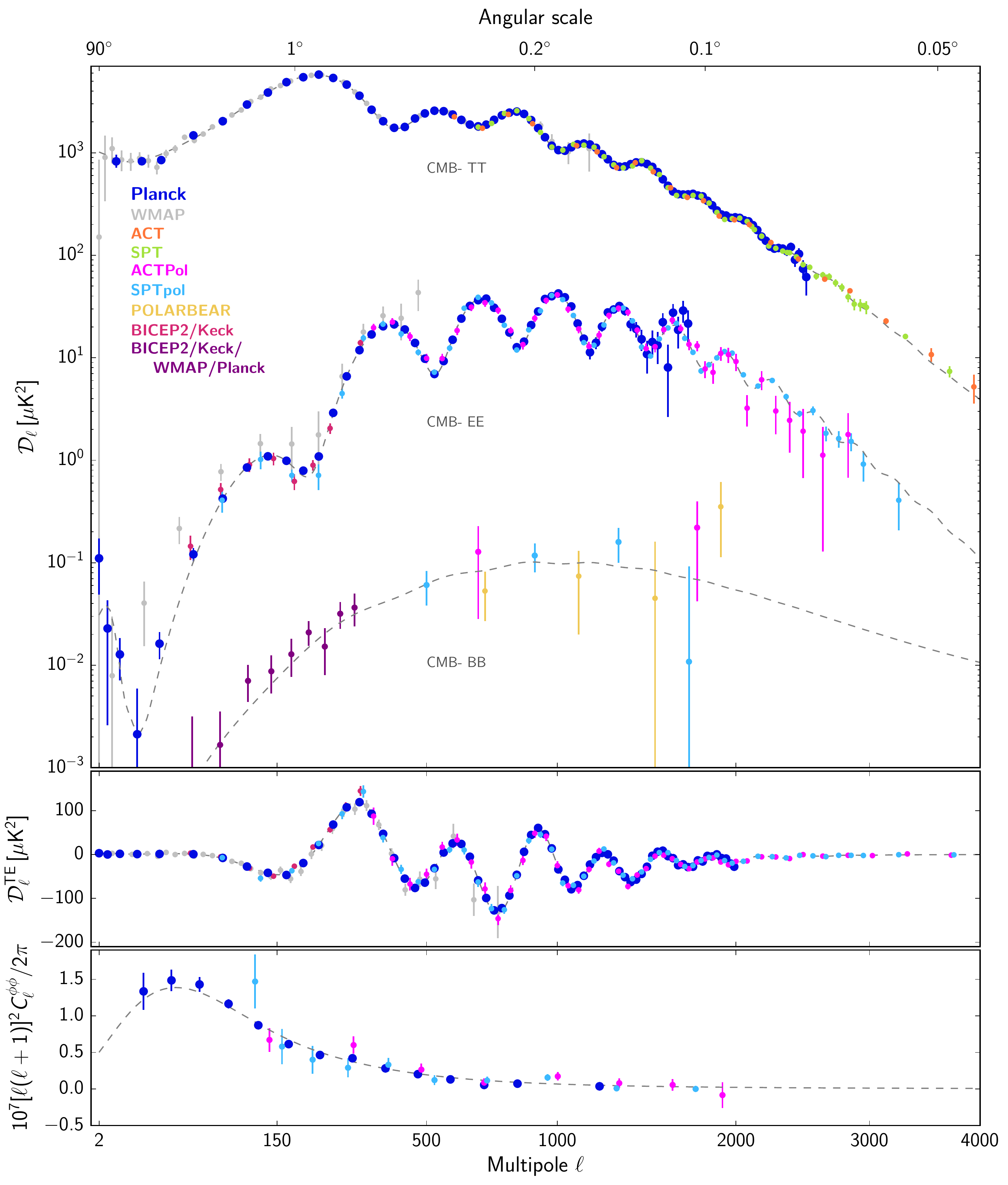}
\caption{CMB power spectra showing the most recent measurements coming from different experiments \cite{planck_18_1}. The top panel shows the auto-power spectra of CMB temperature, E-mode and B-mode anisotropies, while the middle panel shows the TE cross-correlation. The bottom pannel shows the lensing power spectra. The dashed lines show the best-fit $\Lambda$CDM model to Planck temperature, polarisation and lensing data. Planck data cover all the angular scales up to $\ell\approx 2500$, while data from ACT and SPT with finer angular resolution help to reach $\ell\approx 4000$. Figure taken from the Planck Legacy Archive.
}
\label{fig:planck_ps}
\end{figure}

The third generation of CMB space missions, after COBE and WMAP, was ESA's Planck satellite. Planck observed the full sky with an angular resolution of $5$~arcmin \cite{planck_18_1}. In addition to its improved angular resolution and sensitivity, another important improvement of Planck was its wide frequency coverage, with 9 independent frequency channels between 30 and 857 GHz. This allowed a more efficient subtraction of Galactic foregrounds thanks to a more accurate measurement of the spectra of the different emission components, mainly free-free and synchrotron emissions, which are important at low frequencies $\lesssim 50$~GHz, and thermal dust emission, which is important at higher frequencies $\gtrsim 150$~GHz (a concise review on Galactic foregrounds can be found in \cite{dickinson_16}). Figure~\ref{fig:planck_cmb_map} shows the final Planck full-sky map of the primordial CMB anisotropies, resulting from a combination of the different frequency bands to minimise foreground contamination. Grey lines on this map enclose regions with potential residual foreground contamination, mostly along the Galactic plane, that are masked out for cosmological analyses. 

The final Planck power spectrum of these anisotropies ($\mathcal{D}_\ell=\ell(\ell+1)C_\ell/(2\pi)$, where $C_\ell$ is given by equation~\ref{eq:ep_def}), calculated in regions not affected by this mask, is shown by the blue points in Figure~\ref{fig:planck_ps}, in comparison with recent measurements by other experiments. It can be seen that these data allowed the first seven acoustic peaks in the temperature (TT) power spectrum to be clearly measured, and the damping tail of the anisotropies to be traced out to $\ell\approx 2500$. The CMB temperature angular power spectrum is complemented by the polarisation power spectrum, that is usually decomposed into the gradient-like E-mode (even parity) pattern and the curl-like B-mode (odd parity) pattern \cite{zaldarriaga_97,kamionkowski_97}. While E-modes are generated by both scalar and tensor perturbations, B-modes can only be generated by tensor perturbations\footnote{A didactic review on CMB polarisation theory can be found in \cite{hu_97}.} (see Figure~\ref{fig:scalar_tensor_ps}). The EE power spectrum as well as the TE cross-power spectrum have been nowadays measured to high accuracy, while only upper limits on the BB power do exist. The Planck TE and EE power spectra successfully provide high-sensitivity measurements of other 11 peaks (see Figure~\ref{fig:planck_ps}). Planck surpassed all previous experiments in both angular resolution and sensitivity, except for the higher angular resolution measurements of ACT and SPT that allowed the damping tail to be measured out to even higher multipoles.

\begin{figure}
\sidecaption
\includegraphics[scale=1]{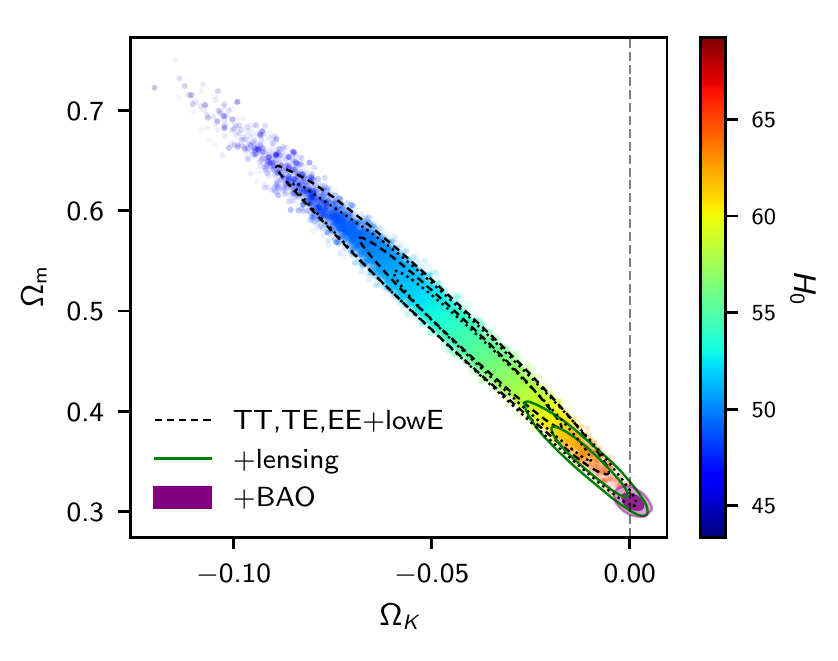}
\caption{Constraints on the $\ok-\om$ plane for a non-flat universe, coming from Planck temperature and polarisation data (dashed and dotted black contours, for two different likelihood implementations), and in combination with Planck lensing data (green contours) and with BAO data (violet regions). In each case 68\% and the 95\% confidence levels are shown. Points show samples from the chains, and their colour indicate the value of the Hubble constant. This shows the well-known ``geometric degeneracy'', which is partially alleviated thanks to Planck lensing data and, more significantly, after the inclusion of BAO data. Figure extracted from \cite{planck_18_6}.
}
\label{fig:planck_ok_om}
\end{figure}

Apart from the improved frequency coverage, sensitivity and angular resolution, Planck data brought two other important improvements over previous missions and experiments. On the one hand, the lower systematics of Planck polarisation data (in particular on the TE cross-correlation and on the EE auto-correlation) for the first time allowed constraints on cosmological parameters to be obtained with comparable accuracies to those derived from temperature data only. On the other hand, Planck data allowed an accurate measurement in the full-sky of the lensing of the CMB photons as they traverse the large-scale structure of the Universe, which essentially leads to a smoothing of the acoustic peaks. This effect was detected for the first time by the ACT telescope \cite{das_11}, and now Planck has measured it with sufficient accuracy as to allow reconstructing a full-sky map of the lensing potential (see Figure~6 of \cite{planck_18_1}), and making it useful for the first time to improve the precision in the determination of cosmological parameters. This also allowed internal parameter degeneracies to be broken in such a way that CMB data can now be used without the need for external cosmological data to derive precise constraints on all 6 $\Lambda$CDM parameters simultaneously. In particular, while the amplitude of the TT power spectrum is proportional to $\as {\rm e}^{-2\tau}$, Planck low-$\ell$ polarisation data allow an independent measurement of $\tau$, therefore breaking the $\as-\tau$ degeneracy associated with temperature data. On the other hand, Planck CMB lensing data contribute to break the well-known CMB ``geometric degeneracy'' in the $\om-\ol$ plane (see Figure~\ref{fig:planck_ok_om}). While previously, the combination of WMAP, ACT and SPT data allowed dark energy to be measured with 3.5\% precision ($\ol=0.721\pm 0.025$ \cite{bennett_13}), now Planck data alone lead to a precision of 0.8\% ($\ol=0.6889\pm 0.0056$ \cite{planck_18_6}). This has contributed to consolidate the CMB as the cosmological probe giving the tightest constraints on cosmological parameters nowadays.

The final Planck constraints on the base 6-parameter $\Lambda$CDM model are shown in Table~\ref{tab:planck_parameters}. The first column corresponds to constraints coming from the TT power spectrum and the low multipoles ($\ell<30$) of the polarisation EE power spectrum, which is used to determine $\tau$ after breaking its degeneracy with $\as$. The results shown in the second and third columns come from the TE and EE power spectra, respectively, while those in the fourth column correspond to a combination of Planck temperature, polarisation and lensing data. The perfect agreement between temperature and polarisation data is a significant achievement of Planck, and a powerful consistency check of the underlying model, and of Planck data themselves. All parameters except $\tau$ are measured with precisions better than $1\%$, with the best-determined parameter being the acoustic scale, $\theta_{\rm MC}$, which is directly related to the Hubble constant, $H_0$, and has precision as good as 0.3\%.

\begin{table}
\caption{Best-fit constraints from CMB-only Planck data \cite{planck_18_6} on the base 6-parameter $\Lambda$CDM model. Different columns correspond to different combinations of temperature, polarisation and lensing CMB data.}
\label{tab:planck_parameters} 
\begin{tabular}{lllllllll}
\hline\noalign{\smallskip}
Parameter &\hspace*{0.2cm}& TT+LowEE &\hspace*{0.2cm}& TE+LowEE &\hspace*{0.2cm}& EE &\hspace*{0.2cm}& TT,TE,EE+lensing\\
\noalign{\smallskip}\svhline\noalign{\smallskip}
$\ob h^2$                           && $0.02212\pm 0.00022$  && $0.02249\pm 0.00025     $  && $0.0240\pm 0.0012  $   && $0.02237\pm 0.00015$ \\
$\oc h^2$                           && $0.1206\pm 0.0021  $    && $0.1177\pm 0.0020     $      && $0.1158\pm 0.0046  $   && $0.1200\pm 0.0012  $ \\
100$\theta_{\rm MC}$        && $1.04077\pm 0.00047$  && $1.04139\pm 0.00049     $  && $1.03999\pm 0.00089$ && $1.04092\pm 0.00031$ \\
$\tau$                                 && $0.0522\pm 0.0080  $    && $0.0496\pm 0.0085     $      && $0.0527\pm 0.0090  $   && $0.0544\pm 0.0073  $ \\
ln(10$^{10}$ A$_{\rm s}$)  && $3.040\pm 0.016    $      && $3.018^{+0.020}_{-0.018}$ && $3.052\pm 0.022    $     && $3.044\pm 0.014  $ \\
$\ns$                                  && $0.9626\pm 0.0057  $    && $0.967\pm 0.011$               && $0.980\pm 0.015    $     && $0.9649\pm 0.0042  $ \\
\noalign{\smallskip}\hline\noalign{\smallskip}
\end{tabular}
\vspace*{-12pt}
\end{table}

While in Table~\ref{tab:planck_parameters} we show the constraints on the 6 independent parameters that form the base $\Lambda$CDM model, Table~2 of \cite{planck_18_1} also lists the constraints on a number of derived parameters. Some of the most interesting ones are the Hubble parameter, $H_0=67.66\pm 0.42$~km~s$^{-1}$~Mpc$^{-1}$, the dark-energy density parameter, $\ol=0.6889\pm 0.0056$, and the age of the Universe, $13.787\pm 0.020$~Gyr. 

Overall the results from Planck data were consistent with previous results from WMAP and other observatories, although with a significant tightening on the precision of most of the measured cosmological parameters. One example is the measured red tilt on $\ns$ with respect to scale invariance ($\ns=1$), which is now measured at $8.4\sigma$, while WMAP measured it at $2.2\sigma$. This is a great success of the $\Lambda$CDM model, but at the same time made feel a bit disappointed people eager to find new issues pointing to new physics beyond $\Lambda$CDM that could have been previously overlooked. There are however some slight differences on some of the previous parameters with respect to WMAP. One important change brought in by Planck data is the measurement of the reionisation optical depth, which is found to be $\tau=0.0544\pm 0.0073$, while WMAP had measured $\tau=0.089\pm 0.014$ \cite{bennett_13}. The smaller value given by Planck is driven by improved cleaning of the Galactic dust emission in polarisation. Actually, when WMAP polarisation maps are cleaned using the Planck 353 GHz channel, they are fully consistent with Planck low-frequency polarisation maps. This is an important achievement of Planck, as it is the fact that the value of $\tau$ coming from Planck low-$\ell$ polarisation data is fully consistent with the value coming from a completely independent dataset as it is the combination of Planck TT, Planck CMB lensing and Baryon Acoustic Oscillation (BAO) data \cite{planck_16_13}. On the other hand, Planck's value of $\ol$ is smaller by $1.3\sigma$ than the one derived by WMAP and, as a consequence, in a flat-geometry $\om$ is larger by $1.3\sigma$. Planck's value of $H_0$ is smaller by $1.0\sigma$, and the age of the universe is bigger by $0.42\sigma$. This is why sometimes it has been said that Planck found that the Universe is slightly fatter and older. The lower value of $H_0$ found by Planck led to some tension with cosmic-distance ladder measurements \cite{riess_19}, at the $\sim 3-4\sigma$ level, and this is nowadays being broadly discussed in the community. This will be commented in more detail in section~\ref{sec:tensions_anomalies}, together with some other possible tensions or anomalies in the data.

The Planck team also explored extensions on the $\Lambda$CDM base 6-parameter model (like the ones commented in section~\ref{sec:extensions}), finding in general good agreement with previous analyses based on other CMB datasets. For instance, no evidence was found of a running of the spectral index (${\rm d}n_{\rm s}/{\rm d ln}k$ in equation~\ref{eq:spectrum_scalar_running}), this being consistent with the predictions of the simplest slow-roll inflation models. When the curvature density parameter is not fixed at $\ok=0$, from temperature and polarisation data it is found $\ok = -0.044^{+0.018}_{-0.015}$, which is an apparent detection of curvature well above $2\sigma$ (see black contours of Figure~\ref{fig:planck_ok_om}). However, the inclusion of CMB lensing data pushes $\ok$ back into consistency with a flat geometry within $2\sigma$ (green contours in Figure~\ref{fig:planck_ok_om}), while including BAO data convincingly breaks the geometric degeneracy (violet regions in Figure~\ref{fig:planck_ok_om}) leading to a perfectly flat geometry with $\ok=0.0007\pm 0.0019$. The low$-\ell$ temperature power spectrum allows the amplitude of tensor modes to be constrained (see black solid line in the right-hand panel of Figure~\ref{fig:scalar_tensor_ps}), giving $r_{0.002}<0.10$ (95\% C.L.) after combining with TE, EE and lensing, at a pivot scale of $0.002$~Mpc$^{-1}$. Combination with polarisation BB power spectra from the BICEP2 and Keck Array experiments \cite{bk_18} of course tightens this constraint to $r_{0.002}<0.065$ (95\% C.L.). This is nowadays the best constraint on a component of tensor modes in the early Universe. Planck data also allowed to study the nature of dark energy by constraining a time-varying equation of state using the parametrisation $w(a)=w_0+(1+a)w_a$ (see section~\ref{sec:dark_energy}). Planck data alone allow a very wide volume of dynamical dark-energy parameter space, although with unrealistically high $H_0$ values. The addition of external data (BAO and SNe) narrows the constraints to the $\Lambda$CDM values, $w_0=-1$, $w_a=0$. In what concerns neutrino masses, while the default Planck analysis assumed the minimal mass $\sum m_\nu=0.06$~eV allowed by neutrino flavour oscillation experiments, leaving it as a free parameter results in a 95\% CL upper limit of $\sum m_\nu<0.54$~eV, when Planck TT and low-$\ell$ EE power spectra are used. Adding Planck polarisation and lensing data tightens this constraint to $\sum m_\nu<0.24$~eV. Interestingly, increasing the neutrino masses leads to lower $H_0$ values, aggravating the tension with the distance-ladder determinations of \cite{riess_19} (see section \ref{sec:tensions_anomalies}). When the number of neutrino relativistic species is left as a free parameter, the resulting constraints are also fully compatible with the standard value of $N_{\rm eff}=3.046$. A detailed discussion on all these analyses, as well as on other extensions to the base $\Lambda$CDM model can be found in \cite{planck_18_6}.

\subsection{The Large Scale Structure of the Universe}
\label{sec:lss}

The density perturbations created during inflation got frozen on the CMB at $z\approx 1100$ (400.000 years after the big bang), and continued evolving to the present day, shaping the large-scale matter distribution visible in the nearby Universe. The three-dimensional spatial distribution of the Large Scale Structure (LSS) of the Universe at low redshifts is very sensitive to several cosmological parameters, and in particular to dark energy. Low-redshift observations of the LSS measure the Universe over the last several billion years, when the dark energy dominates. Comparing the constraints coming from these observations with those derived from the CMB then requires extrapolating predictions to the present-day Universe starting from initial conditions over 13 billion years ago. This is in fact a quite strong test of our cosmological model.

Nowadays the two main LSS observables providing cosmological information are the galaxy clustering and the characterisation of number counts of galaxy clusters.

\subsubsection{Galaxy clustering}
\label{sec:galaxy_clustering}

The same acoustic oscillations in the baryon-fluid plasma in the early Universe that were discussed in section~\ref{sec:cmb}, and which lead to the series of peaks and troughs that are seen in the CMB power spectrum (see Figure~\ref{fig:planck_ps}), create a specific anisotropy in the distribution of galaxies (and of larger structures like clusters of galaxies) in the local Universe. This anisotropy, usually called ``clustering of galaxies'', is mathematically encoded in the two-point correlation function, $\xi(r)$, which gives the expected number of pairs of galaxies with one galaxy in a volume $\delta V_1$ and another galaxy in the volume $\delta V_2$, through the following formula:
\begin{equation}
\langle n_{\rm pair}\rangle = \bar{n}^2\left[1+\xi(r)\right] \delta V_1 \delta V_2~~,
\end{equation}
where $\bar{n}$ is the mean number density of galaxies and $r$ is the separation of the two volume elements. If the galaxies were unclustered then $\xi(r)=0$, so in fact $\xi(r)$ measures the excess clustering of galaxies at separation $r$. It can be shown that $\xi (r)$ is the Fourier transform of the matter power spectrum (see equation~\ref{eq:spectrum_scalar}),
\begin{equation}
\xi (r) = \int P(k) e^{-i{\bf k}.{\bf r}}\frac{d^3 k}{(2\pi)^3}~~.
\end{equation}
Therefore, by measuring the galaxy two-point correlation function we can extract information about the matter power spectrum
\begin{equation}
P(k) = \int \xi (r) e^{i{\bf k}.{\bf r}} d^3 r~~,
\end{equation}
which is the quantity more directly related to theory.

\begin{figure}
\includegraphics[scale=0.5]{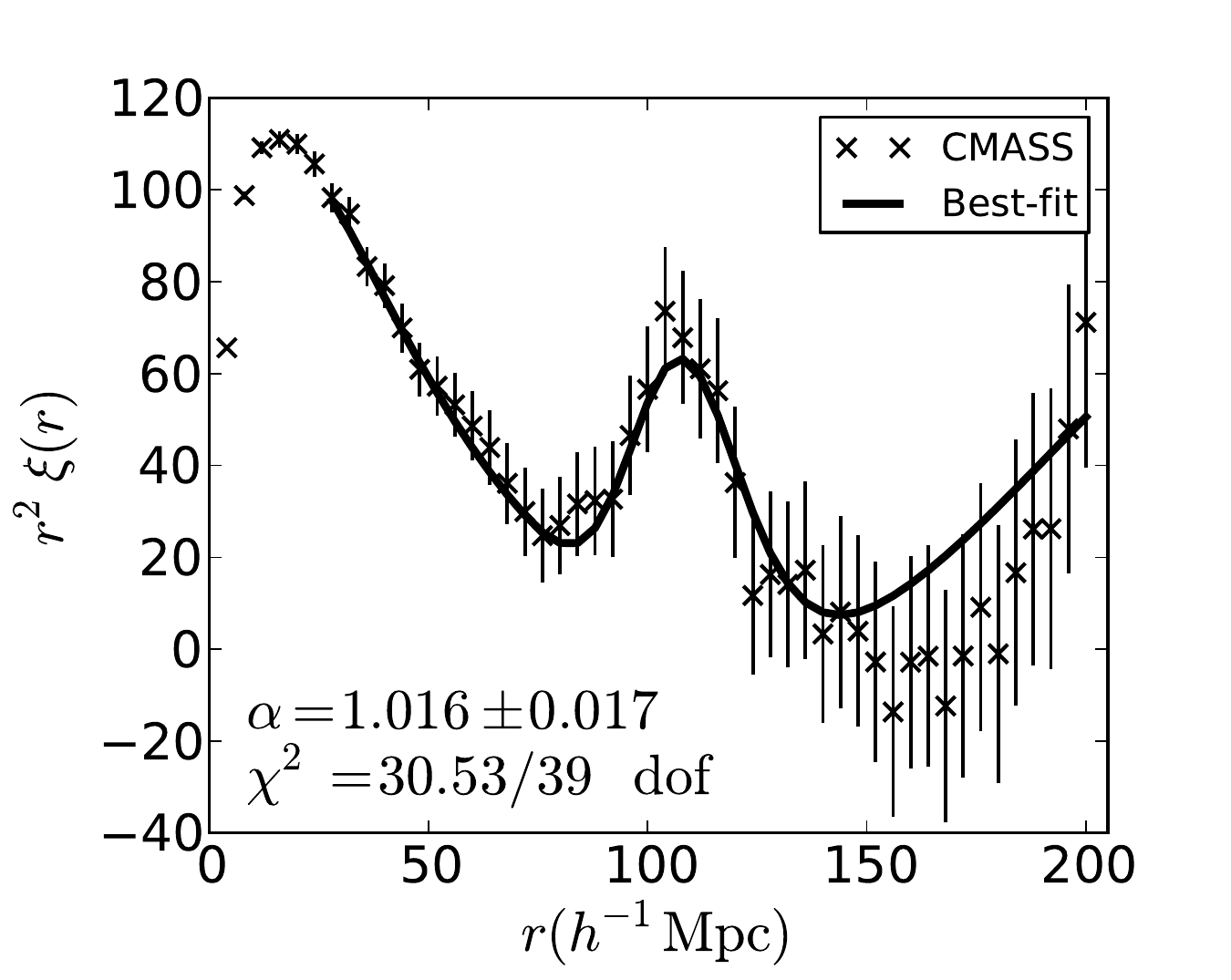}
\includegraphics[scale=0.28]{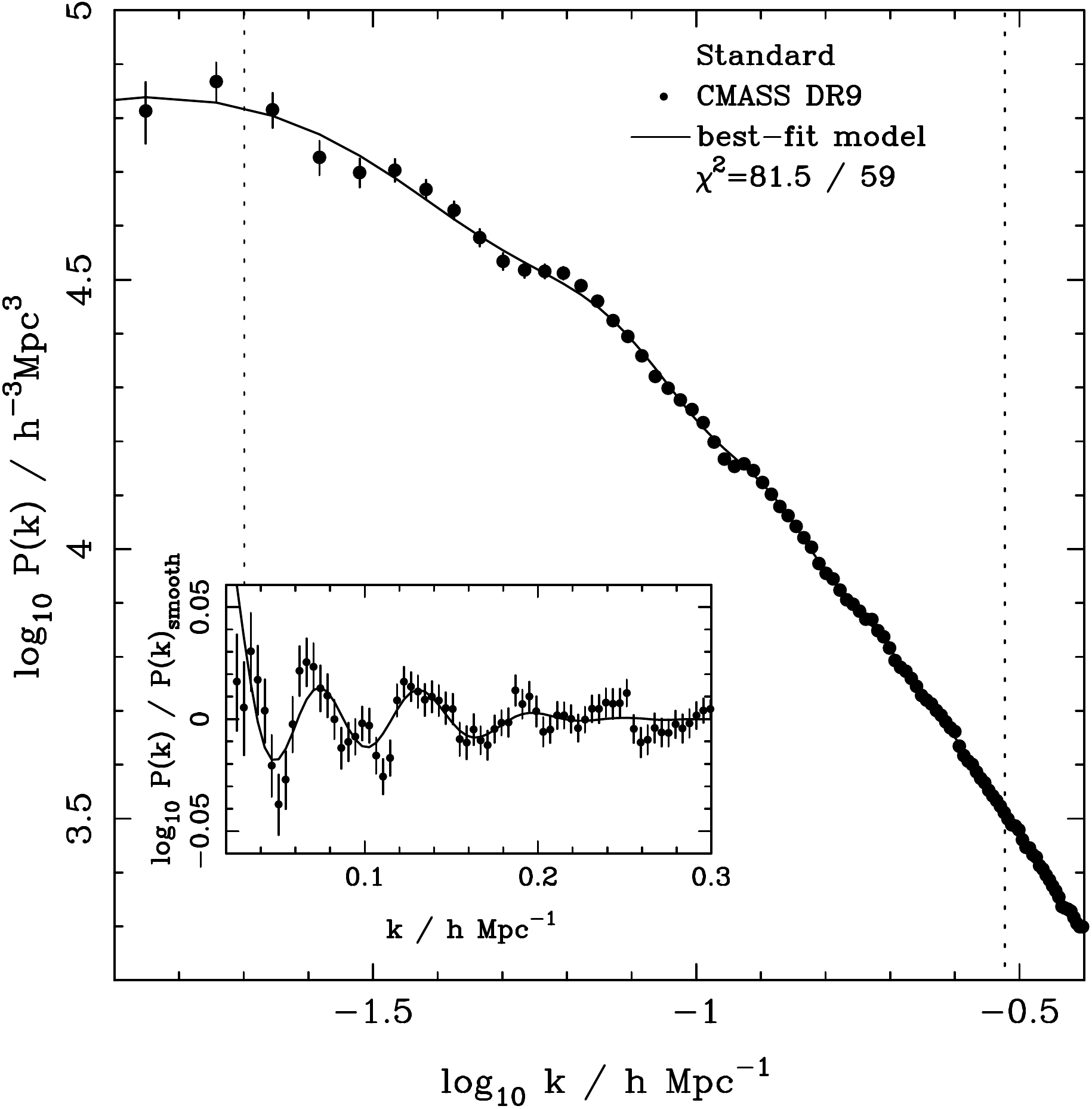}
\caption{LSS correlation function (left) and power spectra (right). Data points come from the BOSS/CMASS galaxy sample, and the solid lines represent the best-fitting models. The left-hand plot shows the best-fitting dilation scale, $\alpha$, which measures the relative position of the BAO peak in the data with respect to a fiducial model, and also the $\chi^2$ statistic giving the goodness of the fit. In the right-hand plot the vertical lines delimit the range of scales ($0.02<k<0.3~h$Mpc$^{-1}$) that are used to fit the data, and the inset shows both the model and the data divided by the best-fitting model with no BAO in the same $k$-range. Figure taken from \cite{anderson_12} (by permission of Oxford University Press).}
\label{fig:cmass_dr9_cf_ps}
\end{figure}

The wiggles visible in the power spectrum (see Figure~\ref{fig:cmass_dr9_cf_ps}) reflect the oscillations of the radiation-baryon plasma, and are coined the Baryon Acoustic Oscillation. They essentially reflect the imprint of the primordial perturbations in the distribution of galaxies in the local Universe, and are equivalent to the acoustic oscillations seen in the CMB power spectrum that were imprinted by the same perturbations at redshift $z\sim 1100$. The BAO wiggles are actually standing sound waves in the pre-recombination universe that imprint a characteristic scale on the late-time matter clustering of galaxies at the radius of the sound horizon,
\begin{equation}
r_d = \int_{z_{\rm d}}^{\infty} \frac{c_s(z)}{H(z)}dz~~,
\end{equation}
evaluated at the drag epoch $z_{\rm d}$, during the decoupling of photons and baryons shortly after recombination. This scale actually provides a ``standard ruler'', that can be measured both in the CMB anisotropies and in the LSS maps at low redshifts. It is seen as a localised peak in the correlation function, or as a series of damped oscillations in the power spectrum (see Figure~\ref{fig:cmass_dr9_cf_ps}). An anisotropic BAO analysis measuring the BAO feature along the line of sight allows the expansion rate $H(z)$ to be constrained, in a highly complementary and independent way to SNe measurements. On the other hand, measuring the BAO feature in the transverse direction allows the comoving angular diameter distance\footnote{For a brief review of cosmology distance definitions see \cite{hogg_00}.} $D_M(z)$ to be constrained. This quantity is related to the physical angular diameter distance by $D_M(z)=(1+z)D_A(z)$. BAO measurements in fact constrain the combinations $H(z)r_d$ and $D_M(z)/r_d$. The cosmological model has then to be adjusted in such a way that radial and angular clustering match, then constraining the product $H(z)D_A(z)$. This was first proposed as a cosmological test by Alcock and Paczynski in 1979 \cite{alcock_79}, and has ever since been coined the Alcock-Paczynski test. If instead we average the clustering in 3D over all directions, then matching the clustering scale to the the expected comoving clustering is sensitive to\footnote{A comprehensive review on galaxy clustering and BAO theory can be found in \cite{percival_13}.}
\begin{equation}
D_V(z) = [c z D_M^2(z)/H(z)]^{1/3}~~.
\label{eq:dv}
\end{equation}

This highlights the importance of mapping the three-dimensional distribution of galaxies in the Universe, in order to be able to reconstruct the matter correlation function and its power spectrum. This requires of course not only measuring the angular positions of galaxies on the sky, which is easy, but also their radial distances, which is not so easy. Although it is possible to obtain photometric redshifts through broad-band colour observations, spectroscopy is clearly the most precise method to obtain redshift estimates. Early spectroscopic galaxy redshift surveys include the Center for Astrophysics Redshift Survey \cite{huchra_83}, the Third Reference Catalogue of Bright Galaxies (RC3; \cite{devaucouleurs_91}) or Las Campanas Redshift Survey (LCRS, \cite{{shectman_96}}). The advent of multi-object spectrographs (MOS) now allows  multiple galaxy spectra to be obtained simultaneously. Key facilities that have used this technique to undertake wide-field redshift surveys include the AA$\Omega$ instrument on the Anglo-Australian Telescope, which has been used to conduct the 2-degree Field Galaxy Redshift Survey (2dF-GRS \cite{{colless_03}}) and WiggleZ \cite{drinkwater_10}, and the Sloan Telescope, which has been used for the Sloan Digital Sky Survey (SDSS; \cite{york_00}). The Baryon Oscillation Spectroscopic Survey (BOSS), part of SDSS-III \cite{eisenstein_11}, was designed to obtain spectroscopic redshifts of $1.5$ million galaxies to at $z=0.2-0.7$ over a sky area of $10,000$ square degrees.

\begin{figure}
\sidecaption
\includegraphics[scale=0.5]{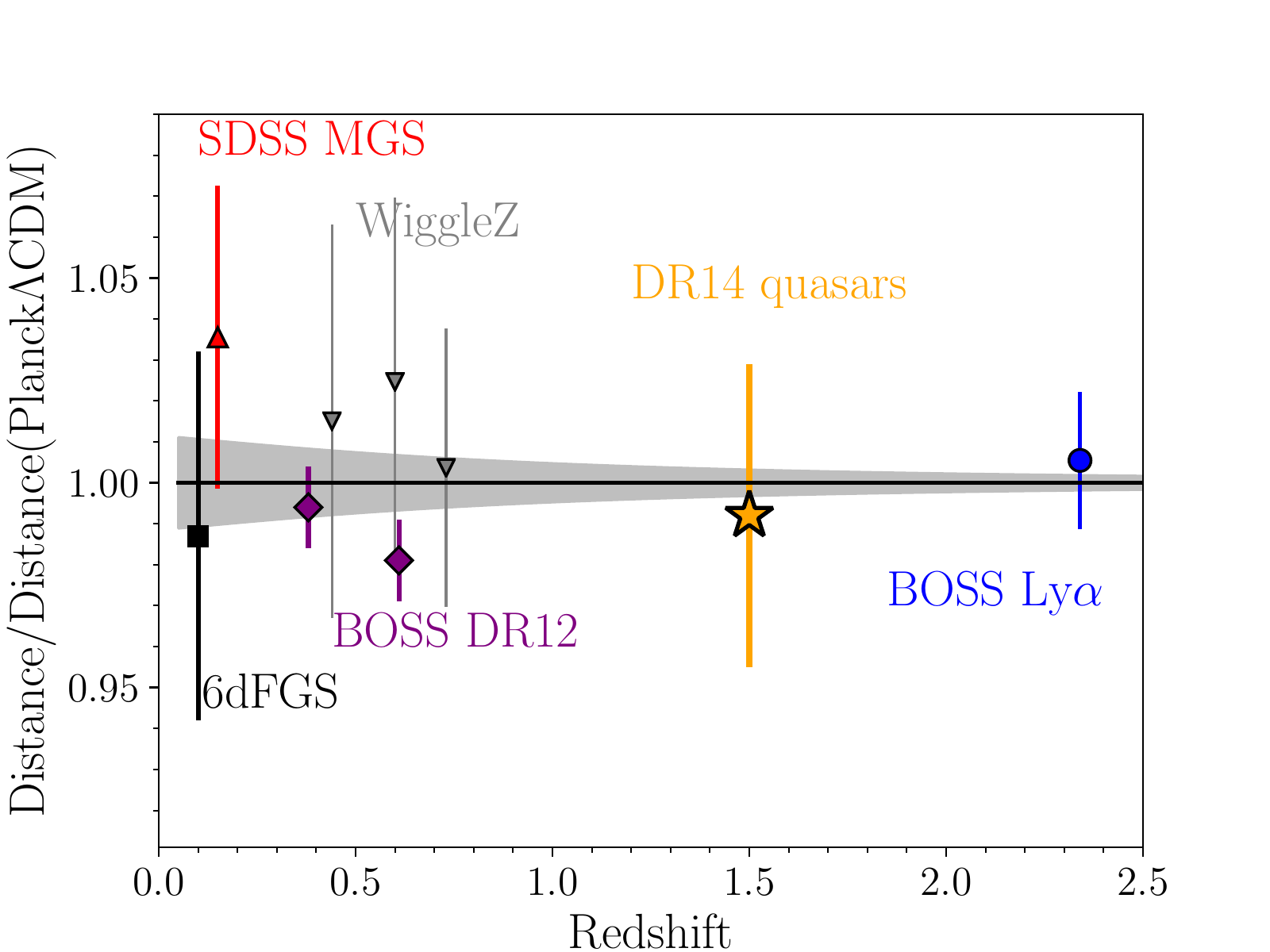}
\caption{BAO distance measurements ($D_V$) coming from BOSS DR12 galaxies \cite{alam_17}, BOSS DR14 QSOs \cite{ata_18}, and from the Ly$\alpha$ forest \cite{bautista_17}, in comparison with measurements from previous surveys at lower redshifts. The results have been normalised to the expected value coming from the best-fit $\Lambda$CDM cosmology from Planck CMB data. The shaded grey area delimits the extrapolated 68\% C.L. region derived from this best-fit CMB model. Figure taken from \cite{ata_18} (by permission of Oxford University Press).}
\label{fig:bao_dist_ata18}
\end{figure}

The first clear detections at low redshift of the BAO signal came from galaxy clustering analyses of the 2dF \cite{cole_05} and from the luminous red galaxy (LRG) sample of the SDSS \cite{eisenstein_05}, at redshift $z\approx 0.3$. The WiggleZ survey allowed the redshift coverage to be increased to $0.4<z<1.0$, with a precision of 3.8\% \cite{drinkwater_10}. The initial goals of the BOSS survey were to improve the precision of the measurement down to 1\% in the redshift range $z=0.2-0.7$, and to achieve for the first time a detection of the BAO peak at $z>2$ using the 3D structure in the Ly$\alpha$ forest absorption towards $160,000$ high-redshift quasars. Both goals were successfully fulfilled. Figure~\ref{fig:cmass_dr9_cf_ps} shows the galaxy correlation function $\xi(r)$ and power spectrum $P(k)$ derived from the CMASS sample of the BOSS survey \cite{anderson_12}, corresponding to data release 9 (DR9). The CMASS DR9 samples comprises 264,283 galaxies distributed over 3275 square degrees, and with redshifts $0.43<z<0.7$. Further data releases have resulted in an increased number of galaxies distributed over progressively larger sky areas. A catalogue of positions and redshifts of $147,000$ QSOs at $0.8<z<2.2$ extracted from the extended BOSS (eBOSS) data \cite{ata_18}, as well as the absorption by the Ly$\alpha$ forest towards the line of sight of $157,783$ QSOs at $2.1<z<3.5$ \cite{bautista_17}, have also been used to reconstruct the BAO signature. Figure~\ref{fig:bao_dist_ata18} shows a summary of the BAO distance measurements ($D_V$, which is a combination of $H(z)$ and $D_M(z)$, see equation~\ref{eq:dv}) resulting from these measurements, as well as from previous surveys at lower-redshifts, normalised by the expected value coming from the best-fit $\Lambda$CDM model to the Planck CMB observations. The excellent agreement between independent BAO measurements at different redshifts, as well as with the CMB measurements at $z=1100$, is a remarkable success of both the observations and the theoretical model.

\begin{figure}
\sidecaption
\includegraphics[height=3.8cm]{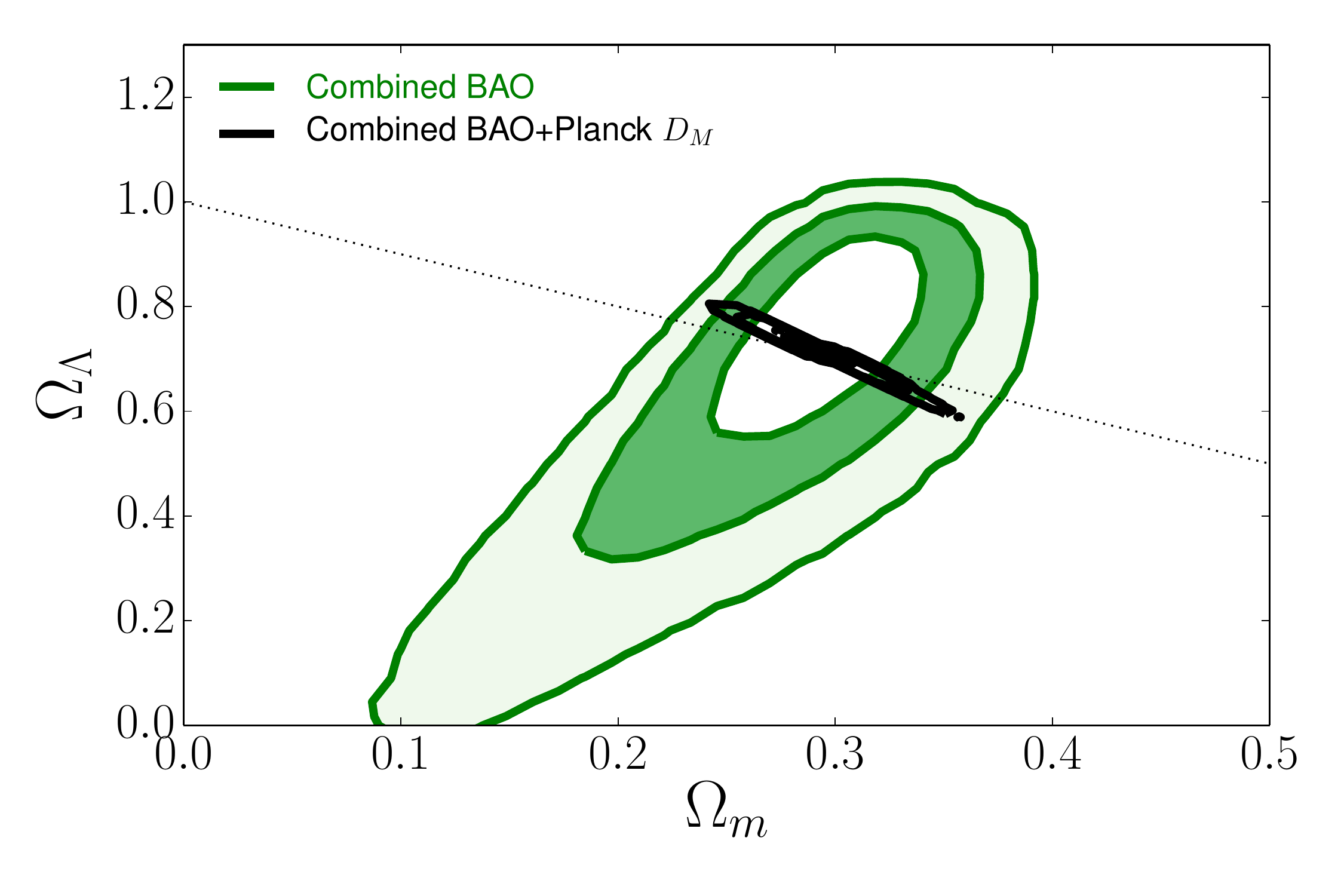}
\includegraphics[height=3.8cm]{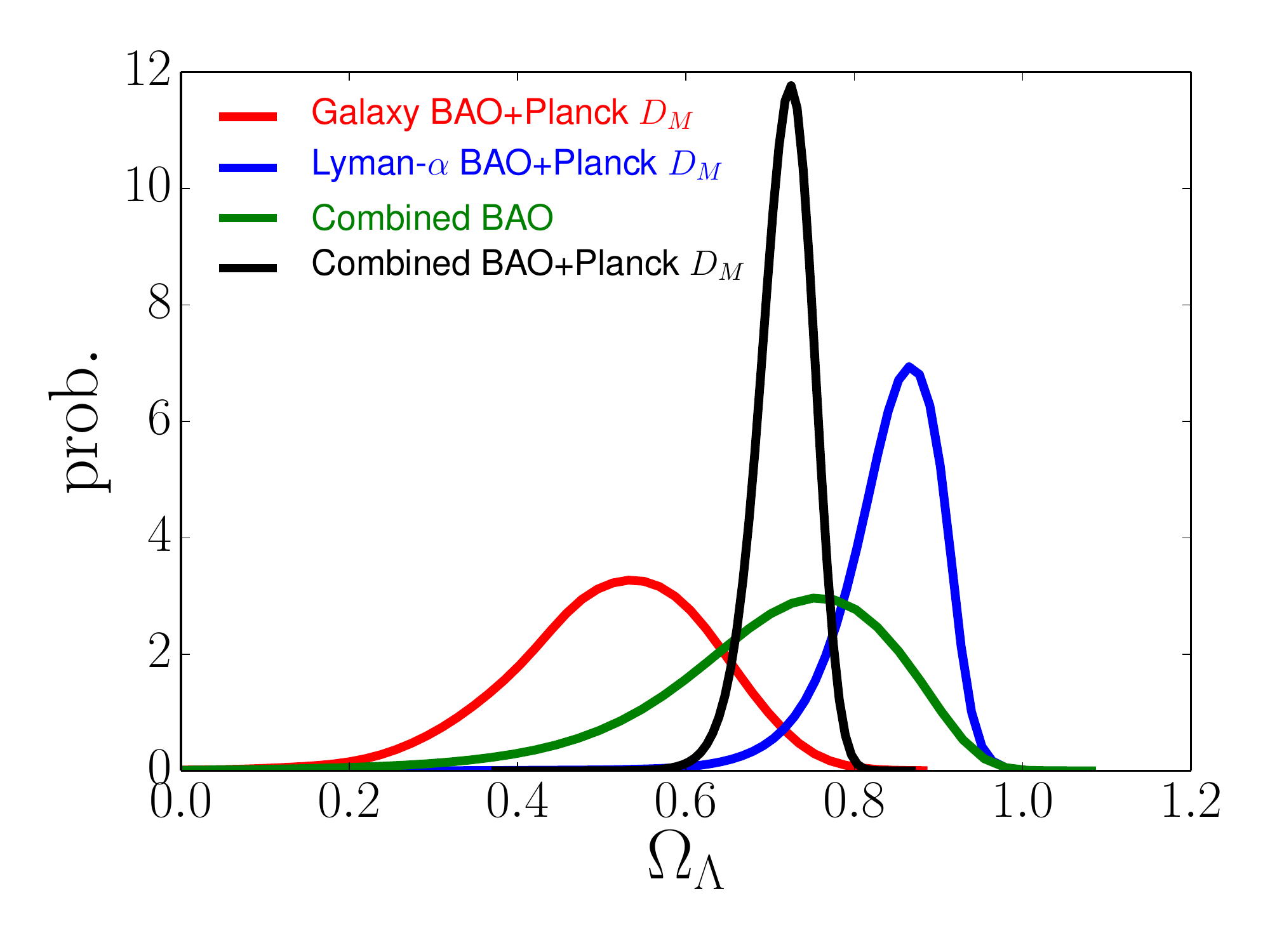}
\caption{Constraints derived from BOSS BAO measurements on open $\Lambda$CDM models. The left panel shows constraints on the $\ol-\om$ plane (dotted line corresponds to a flat geometry), while the right panel shows the $\ol$ one-dimensional probability density function. Green contours and lines show constraints using only BAO measurements, derived from the combination of galaxy and Ly$\alpha$ forest measurements, while black represents the combination of BAO and Planck CMB measurements. The results derived from the combination of the CMB with either BAO galaxy measurements or BAO Ly$\alpha$ measurements are shown in red and blue, respectively. Figure taken from \cite{aubourg_15}.}
\label{fig:ol_om_bao}
\end{figure}

As it was explained before, these BAO measurements are crucial to set constraints on cosmological parameters, in particular on the expansion rate and thus on dark energy. Also, combination with the CMB crucially breaks some parameter degeneracies, allowing for even tighter cosmological constraints (see Figure~\ref{fig:planck_ok_om}). As an example, in Figure~\ref{fig:ol_om_bao}, we show constraints on $\om$ and $\ol$ coming from BAO measurements alone, and from the combination of BAO and CMB. Nowadays BAO data alone can actually allow dark energy to be measured at more than $3\sigma$, while combination with CMB measurements from Planck increases the significance to $21\sigma$ \cite{aubourg_15}.

\subsubsection{Galaxy clusters}
\label{sec:galaxy_clusters}

Different observables related to galaxy clusters have important applications in cosmology. The clustering of galaxy clusters, and their BAO signature, can be used to constrain cosmological parameters in the same way as the clustering of galaxies \cite{marulli_18}, although typically with lower constraining power due to the smaller number of objects. The combination of X-ray and Sunyaev-Zel'dovich (SZ) effect (this is the inverse Compton scattering of CMB photons as they transverse the hot gas inside galaxy clusters) observations can be directly used to mesure cosmic distances, and in turn to estimate the Hubble constant at the redshift of the cluster \cite{bonamente_06}, although this method is prone to different kind of systematic effects and is not very popular nowadays. Peculiar velocities of galaxy clusters, which can be measured through the kinetic SZ effect, probe directly the gravitational potential. Estimates of the gas mass fraction through X-rays or through the SZ effect can be used to infer the matter density parameter, $\om$, if the baryonic density $\ob$ is fixed from any other method. The power spectrum of the temperature anisotropies induced by the SZ effect can be used to constrain a combination of $\om$ and of the amplitude of matter fluctuations on scales of $8~h^{-1}$~Mpc, $\sigma_8$; more specifically it allows the product $\sigma_8^8 \om^3$ to be constrained \cite{planck_16_22}.

\begin{figure}
\sidecaption
\includegraphics[height=5.5cm]{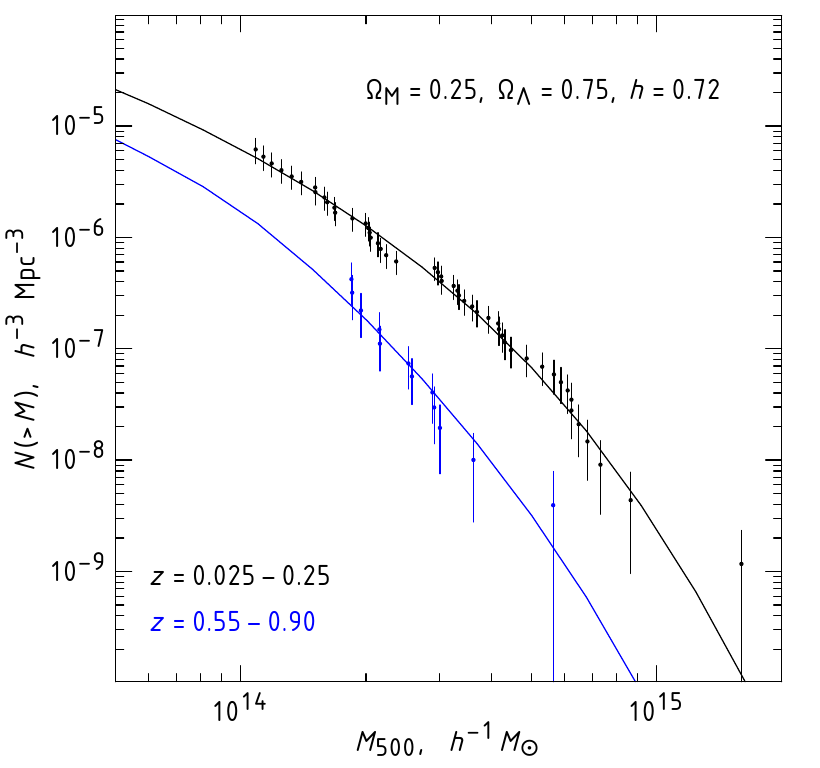}
\includegraphics[height=5.5cm]{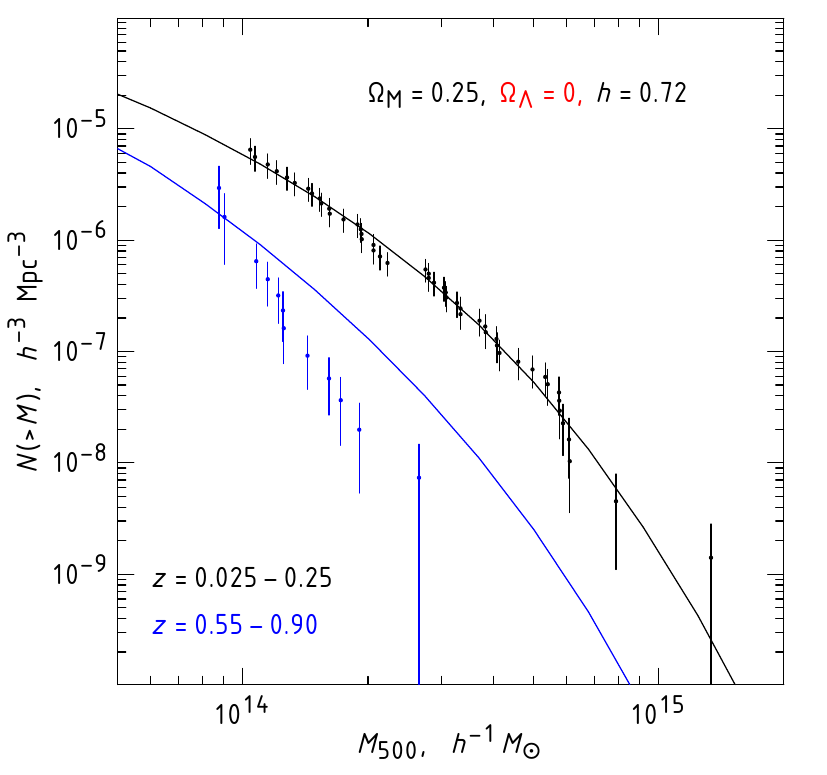}
\caption{Illustration of the sensitivity of the mass function of clusters of galaxies to the cosmological model. The data and the model have been computed for a model with (left panel) and without (right panel) dark energy, considering a low- and a high-redshift galaxy sample. It becomes immediately clear that a model without dark energy provides a very poor fit to the data in the high redshift bin. Figure taken from \cite{vikhlinin_09} ($\copyright$ AAS. Reproduced with permission).}
\label{fig:ol_clusters}
\end{figure}

Apart from all these observables and applications, the most popular application of galaxy clusters to cosmology nowadays comes from the characterisation of their number counts, which essentially gives the number of clusters per redshift and mass bin, in a given solid angle element $\Delta\Omega_i$:
\begin{equation}
\bar{N}(z_i,M_j) = \frac{\Delta\Omega_i}{4\pi}\int_{z_i}^{z_{i+1}} dz \frac{dV}{dz} \int_{{\rm ln}M_j}^{{\rm ln}M_{j+1}} d{\rm ln}M \frac{dn}{d{\rm ln}M}~~.
\end{equation}
Cosmology enters here through the volume element $dV/dz$, and through the mass function, $dn/d{\rm ln}M$. While the amplitude of the mass function is proportional to $\sigma_8$, its shape is proportional to $\om$. There is a degeneracy between these two parameters, in such a way that these analyses constrain the product $\sigma_8\om^{0.3}$. This degeneracy is broken by using external priors on other parameters. In addition to the shape of the local mass function, its redshift evolution, which is sensitive to the growth of linear density perturbations, can be used to constrain the dark energy density parameter and its equation of state. This was done by \cite{vikhlinin_09} using a sample of just 86 galaxy clusters, from which they obtained a $5\sigma$  detection of dark energy (see Figure~\ref{fig:ol_clusters}).

Obtaining meaningful cosmological constraints from the galaxy clusters mass function requieres statistically-large cluster samples together with accurate estimates of redshifts and masses. While nowadays X-ray and SZ surveys provide large samples and redshifts can be estimated to high accuracy by follow-up spectroscopic observations, the mass estimates are subject to systematics that are likely to introduce biases on the inferred cosmological parameters. Different mass proxies can be used, like the X-ray luminosity, the SZ flux, the optical richness, the velocity dispersion of individual galaxies, or weak lensing measurements. Of course, a good proxy should have a small scatter between the estimated and the real cluster mass, and a low bias. All cosmological analyses assume a certain bias between the estimated and the real mass, which must be fixed or fitted for, and is denoted by $(1-b)$. This bias can result either from cluster physics that are not properly accounted for or from instrumental effects. 

\begin{figure}
\sidecaption
\includegraphics[height=5.8cm]{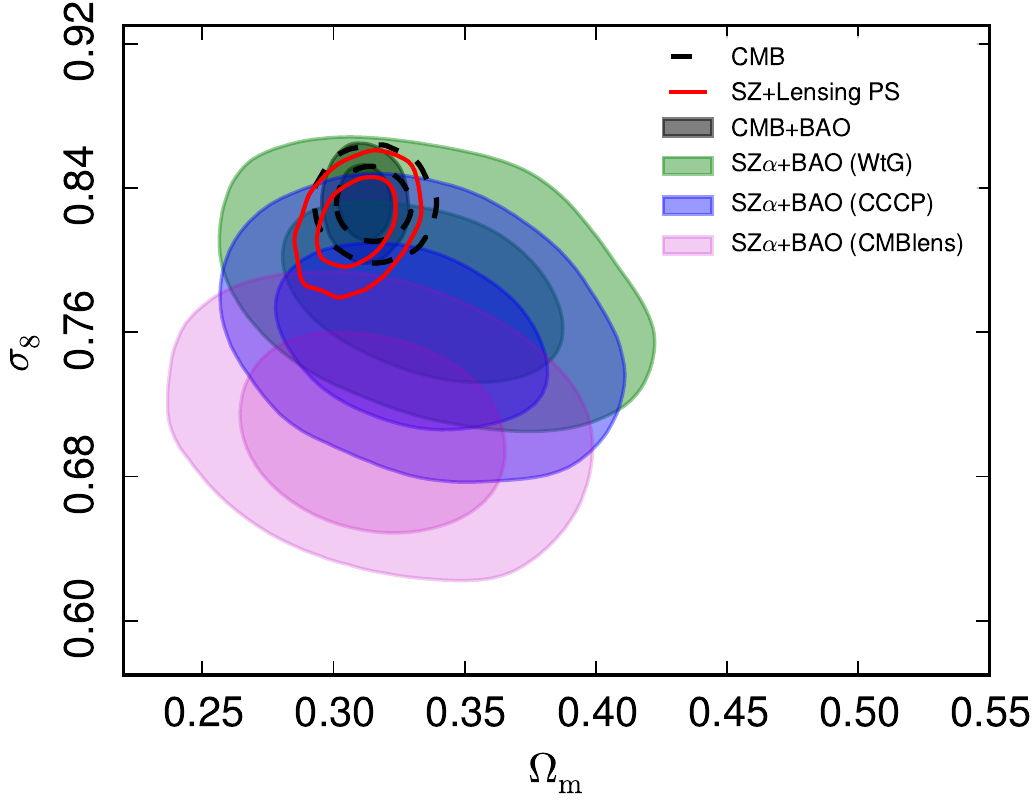}
\caption{Constraints on the $\om-\sigma_8$ plane derived from SZ cluster number counts by the Planck collaboration \cite{planck_16_24}, using different mass estimates (WtG, CCCP and CMBlens), compared with those coming from the primary CMB (black contours), from the primary CMB and BAO (black shaded area) and from the combination of primary CMB and the lensing power spectrum (red contours). Credit: \cite{planck_16_24}, reproduced with permission $\copyright$ ESO.}
\label{fig:om_s8_clusters}
\end{figure}

Thanks to its large and well characterised catalogue of galaxy clusters detected via the SZ effect, the results from the Planck collaboration led to an important step forward in this kind of analyses. The 2013 Planck cosmological analysis using cluster number counts \cite{planck_13_20} relied on the $Y_{\rm X}-M_{500}$ relation (relation between X-ray luminosity and the mass enclosed inside $r_{500}$, the radius at which the density is 500 times the critical density of the Universe) as a mass proxy, and either fixed the mass bias parameter at $(1-b)=0.8$, or used a flat prior inside the range $[0.7,1.0]$. On the other hand, the 2015 Planck analysis \cite{planck_16_24} relied on improved mass proxies based on weak lensing measurements, and also on the use of the lensing of the CMB photons on the cluster position, a new technique that is starting to be exploited. Using two different cluster samples with measurements of the gravitational shear they obtained respectively $(1-b)=0.688\pm 0.072$ (WtG in Figure~\ref{fig:om_s8_clusters}) and $(1-b)=0.780\pm 0.092$ (CCCP in Figure~\ref{fig:om_s8_clusters}), while the CMB lensing measurement led to $(1-b)=1.01\pm 0.19$. The scatter between these three estimates clearly highlights the importance of having a reliable mass estimate in order for this method to produce accurate cosmological constraints, in particular on the $\sigma_8$ parameter, which is the one more strongly affected by this issue. Figure~\ref{fig:om_s8_clusters} shows the final constraints on the $\om-\sigma_8$ plane obtained by the Planck collaboration from the 2015 results (based on DR2), and using their three mass estimates. It can be seen in this figure that while the WtG method leads to agreement with the CMB, the $\sigma_8$ estimates derived from the other two methods are in tension with the CMB. The value of the mass bias that is required to reconcile both measurements is $(1-b) = 0.58\pm 0.04$, which is close to the first of the two previous estimates. 

\subsection{Type Ia Supernovae}
\label{sec:sn}
Type Ia supernova (hereafter SNIa) are thought to arise from thermonuclear explosions of white dwarfs in binary systems, and are known to produce rather uniform brightness at their maximum, therefore being considered as standard candles as their visual magnitude depends primarily on their distance to the observer. Their V-band peak luminosities present a scatter of approximately 0.4~mag \cite{riess_96}. This scatter can be notably reduced by introducing some corrections, the most important of which has to do with the correlation between the peak luminosity and the light curve shape (LCS), first introduced by \cite{phillips_93}. Two other important refinements are the correction for the correlation between SN colour and extinction and  the application of $K$-corrections for the redshifting effects. After these corrections are introduced, the dispersion in well-measured optical band peak magnitudes is reduced to only $\sim 0.12$~magnitudes, allowing each measured SN to provide a luminosity-distance estimate with precision of $\sim 6$\%.

\begin{figure}
\includegraphics[height=4.7cm]{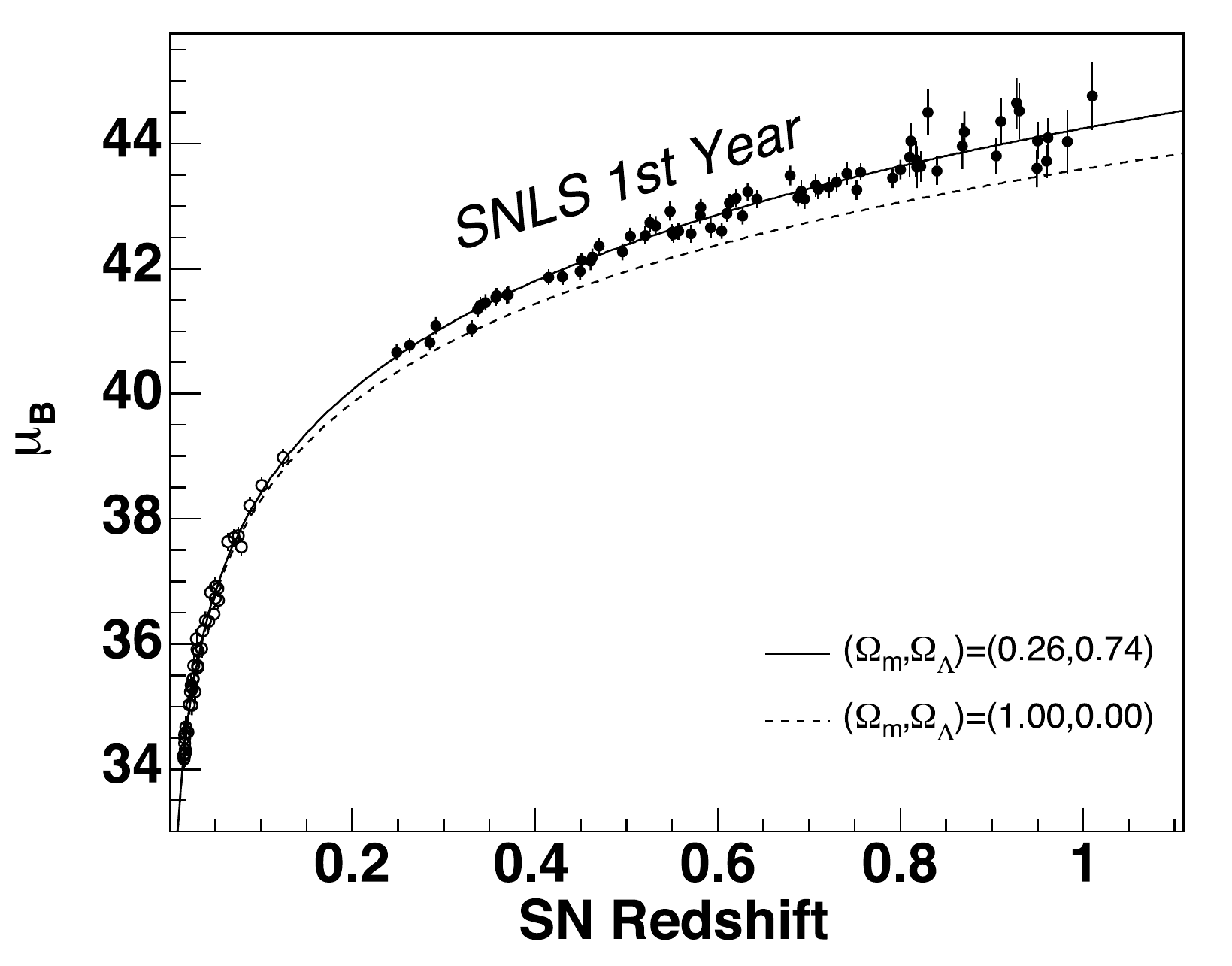}
\includegraphics[height=4.7cm]{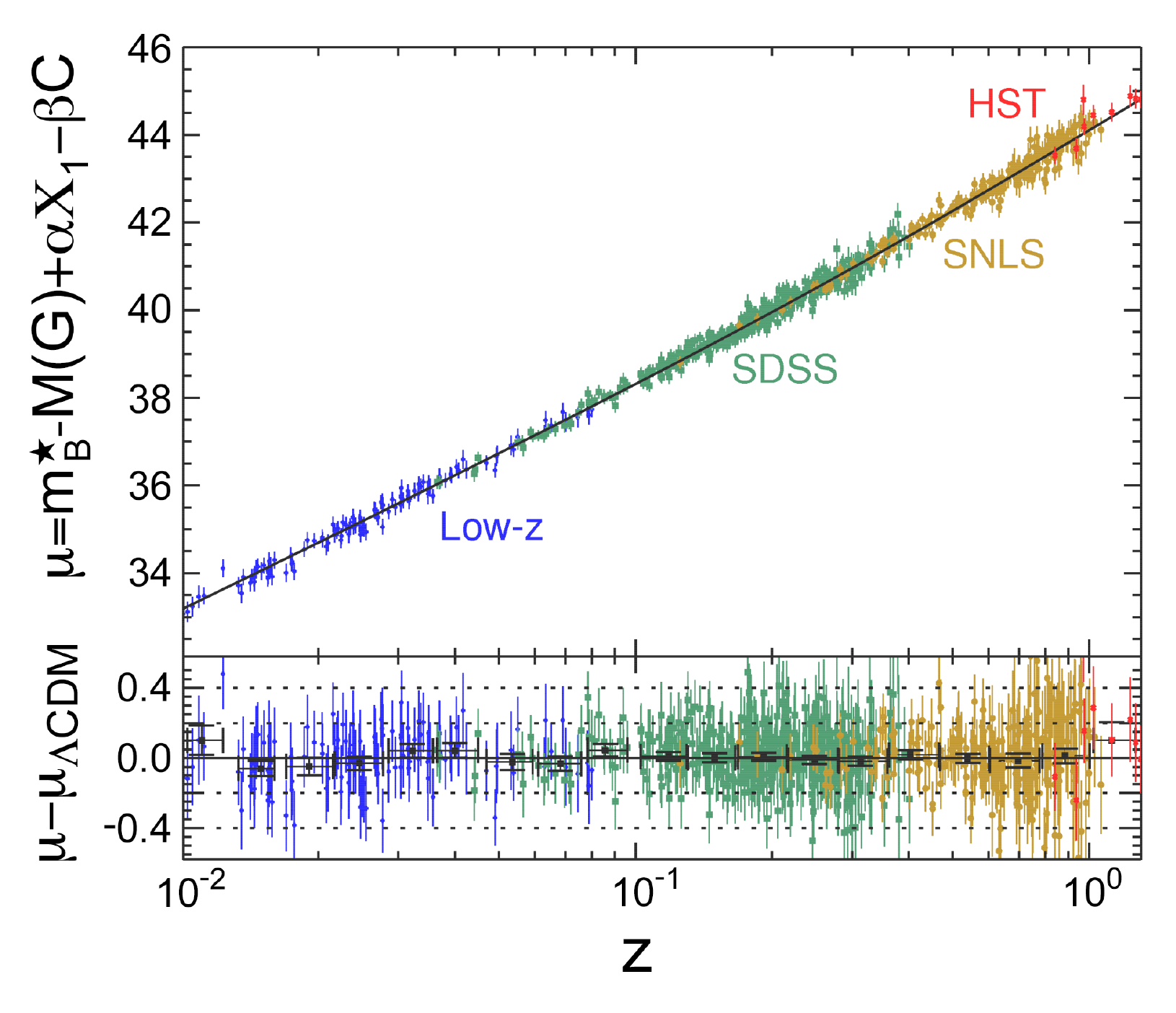}
\caption{Hubble diagrams showing data points obtained from Type Ia supernova observations. The left panel shows measurements from the Supernova Legacy Survey ($z>0.2$) in combination with nearby SNIa measurements. Two cosmological models are shown superimposed, one corresponding to the best-fit cosmological model with $\ol=0.74$, and another one without dark energy, clearly highlighting that the data favour $\ol>0$. The right panel shows the results from the joint light-curve analysis (JLA), with combined measurements from four different surveys, providing uniform redshift coverage out to $z=1$, together with the distance modulus redshift relation of the best-fit $\Lambda$CDM cosmology. Credit: \cite{astier_06} (left) and \cite{betoule_14} (right), Reproduced with permission $\copyright$ ESO.}
\label{fig:hubble_diagram}
\end{figure}

Their ability to measure distances to such a good precision bestows SNIa important cosmological applications, as the luminosity-distance is sensitive to the density parameters $\om$, $\ok$, $\ol$ and to the expansion rate $H(z)$ (see relevant equations in \cite{hogg_00}). At low redshifts, where the cosmic expansion is still linear, and therefore the effects of curvature and dark-energy are negligible (see Figure~\ref{fig:hubble_diagram}), SNIa observations can be used to measure this linearity and thence to obtain estimates of the Hubble constant \cite{sandage_82}. This requires knowledge of the absolute SNIa magnitude, which is usually calibrated through cepheid distances, using the period-luminosity relation (Leavitt law). The most-commonly used anchors are geometric distances to Milky Way Cepheids, eclipsing binaries in the Large Magellanic Clouds or in M31, and the water megamasers in NGC4258. This was the approach that was followed by the HST Key Project, to get an estimate with 11\% precision, $H_0=72\pm 8$~km~s$^{-1}$~Mpc$^{-1}$ \cite{freedman_01}. The error of this measurement is driven by systematics related to the absolute calibration, and therefore recent efforts have focused on improving this calibration rather than on decreasing the statistical errors by increasing the number of objects. This has been the goal of the SH0ES (``Supernova and $H_0$ for the Equation of State'') programme, whose most recent measurement is $H_0=74.03\pm 1.42$~km~s$^{-1}$~Mpc$^{-1}$, which corresponds to $1.9\%$ precision \cite{riess_19} and is in tension at a level of $4.3\sigma$ with the CMB measurement from Planck (see section~\ref{sec:cmb}, and further discussion in section~\ref{sec:tensions}). Recently, the ``Carnegie-Chicago Hubble Program'' (CCHP), using a completely independent calibration based on the tip of the red giant branch (TRGB) method, which uses population II stars, derived a considerably lower value, $H_0=69.8\pm 1.9$~km~s$^{-1}$~Mpc$^{-1}$ \cite{freedman_19}, which sits in the middle between the SH0ES programme and the CMB values. Also recently, \cite{macaulay_18} obtained $H_0=67.77\pm 1.30$~km~s$^{-1}$~Mpc$^{-1}$ after calibrating the intrinsic magnitudes of 207 SNIa measured by the ``Dark Energy Survey'' (DES) at $0.018<z<0.85$, plus other 122 SNIa at lower redshifts, using the ``inverse distance ladder'' method, which uses BAO measurements as a reference. This method requires fixing the sound horizon scale, for which they used the value measured by Planck. While this measurement derived from SNIa is closer to the CMB value, it has to be taken into account that it is not fully independent of the CMB measurements. 

At cosmologically-significant distances, where the effects of the matter and density content of the Universe become important, the luminosity distance is obtained through an integration of the expansion rate over the redshift, which depends on the cosmological model. Therefore, the dimming of the standard candles at high redshifts can be used to constrain cosmological parameters. Importantly, contrary to the local measurement of $H_0$, the measurement of the expansion rate at high redshifts is independent of the absolute luminosity of the SNIa, which however is considered to be constant with redshift. In the late 90s two independent teams, the ``High-z Supernova Research Team'' \cite{riess_98} and the ``Supernova Cosmology Project'' \cite{perlmutter_99}, unexpectedly found that distant supernova (out to $z=0.8$) are $\sim 0.25$~mag dimmer than they would be in a decelerating universe, indicating that the Universe is currently undergoing an accelerated expansion. When analysed assuming a universe with matter and a cosmological constant, these results provided evidence of $\ol>0$ at more than 99\%. This important discovery was recognised with the 2011 Physics Nobel Prize. Since then, new observations have compellingly confirmed this result. The largest high-redshift ($z\approx 0.4-1.0$) samples to date come from the ESSENCE survey \cite{wood-vasey_07} and from the CFHT Supernova Legacy Survey (SNLS \cite{astier_06}). The left-hand panel of Figure~\ref{fig:hubble_diagram} shows the Hubble diagram resulting from the SNLS high-z SNIa sample, in combination with other observations at lower redshifts. Following what has become common practice in this field, instead of distances this figure represents magnitudes versus redshift. It is clearly seen that a model with $\ol=0$ is strongly disfavoured by the data. At intermediate distances ($0.1<z<0.4$) the SDSS-II supernova survey \cite{sako_18} has resulted in 500 spectroscopically confirmed SNIa. At very high redshifts, HST surveys (see e.g. \cite{riess_07}) have yielded $\sim 25$ SNIa at $z>1$ suitable for cosmological analyses. 

The greatest cosmological utility of SNIa comes from the combination of different data sets spanning a wide redshift range. However, combination and homogenisation  of data from different instruments and telescopes in such a way that they are useful for cosmological analyses entails a major difficulty, specially in what concerns survey-to-survey relative flux calibration, joint light curve fitting, and consistent use of $K$-corrections. The joint light-curve analysis (JLA; \cite{betoule_14}) contains 740 spectroscopically confirmed SNIa with high-quality light curves, coming from the combination of two major surveys, the SDSS-II supernova survey and the SNLS, in addition to low-$z$ observations, and very high-$z$ data from the HST (see right-hand panel of Figure~\ref{fig:hubble_diagram}). This was the default sample for cosmological analyses, until the recent advent of the ``Pantheon'' sample \cite{scolnic_18}, which contains 1048 SNe spanning the redshift range $0.01<z<2.3$, thanks to the addition of Pan-STARRS1 Medium Deep Survey and various low-redshift and HST samples.

\begin{figure}
\includegraphics[height=5.7cm]{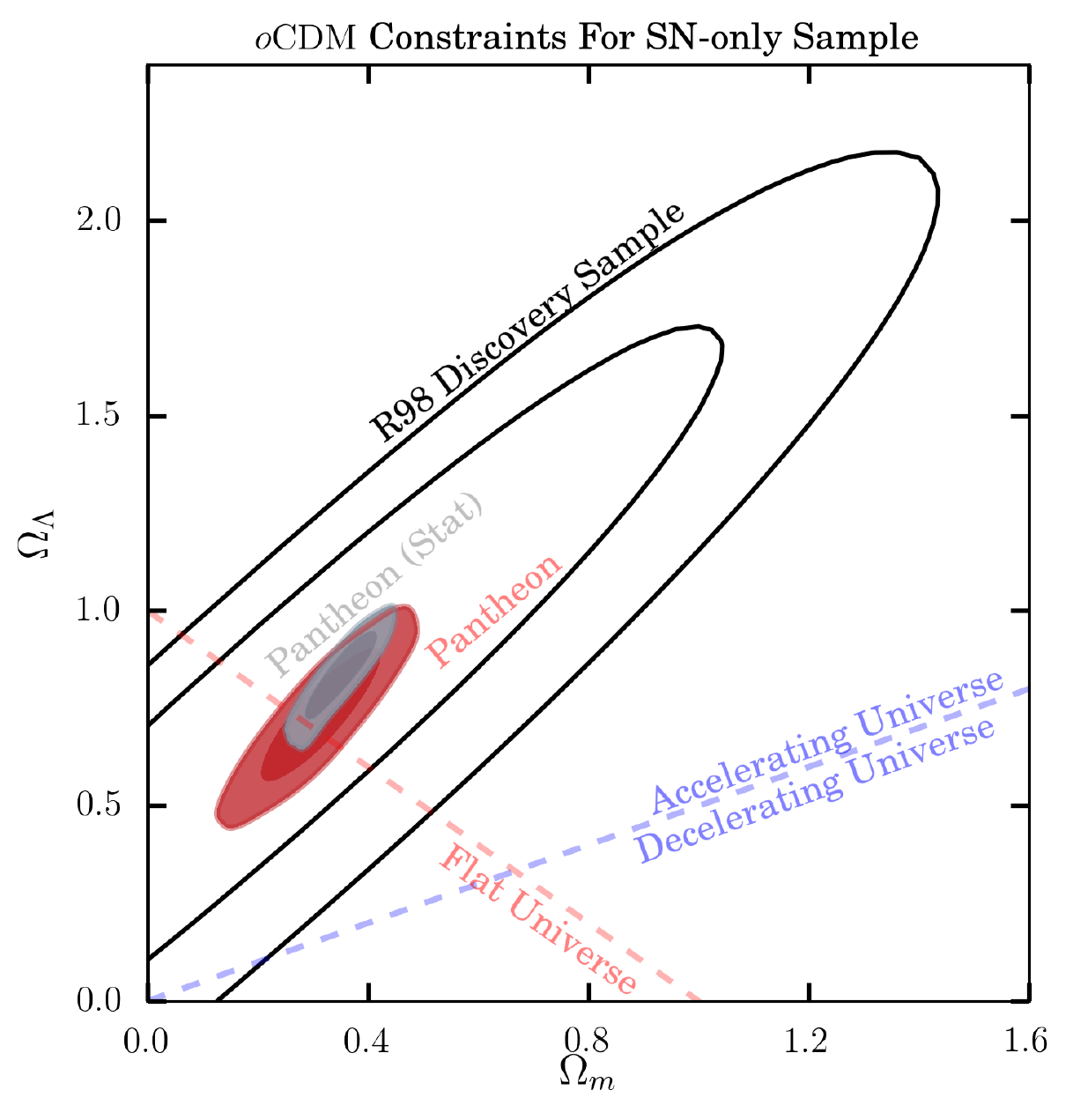}
\includegraphics[height=5.7cm]{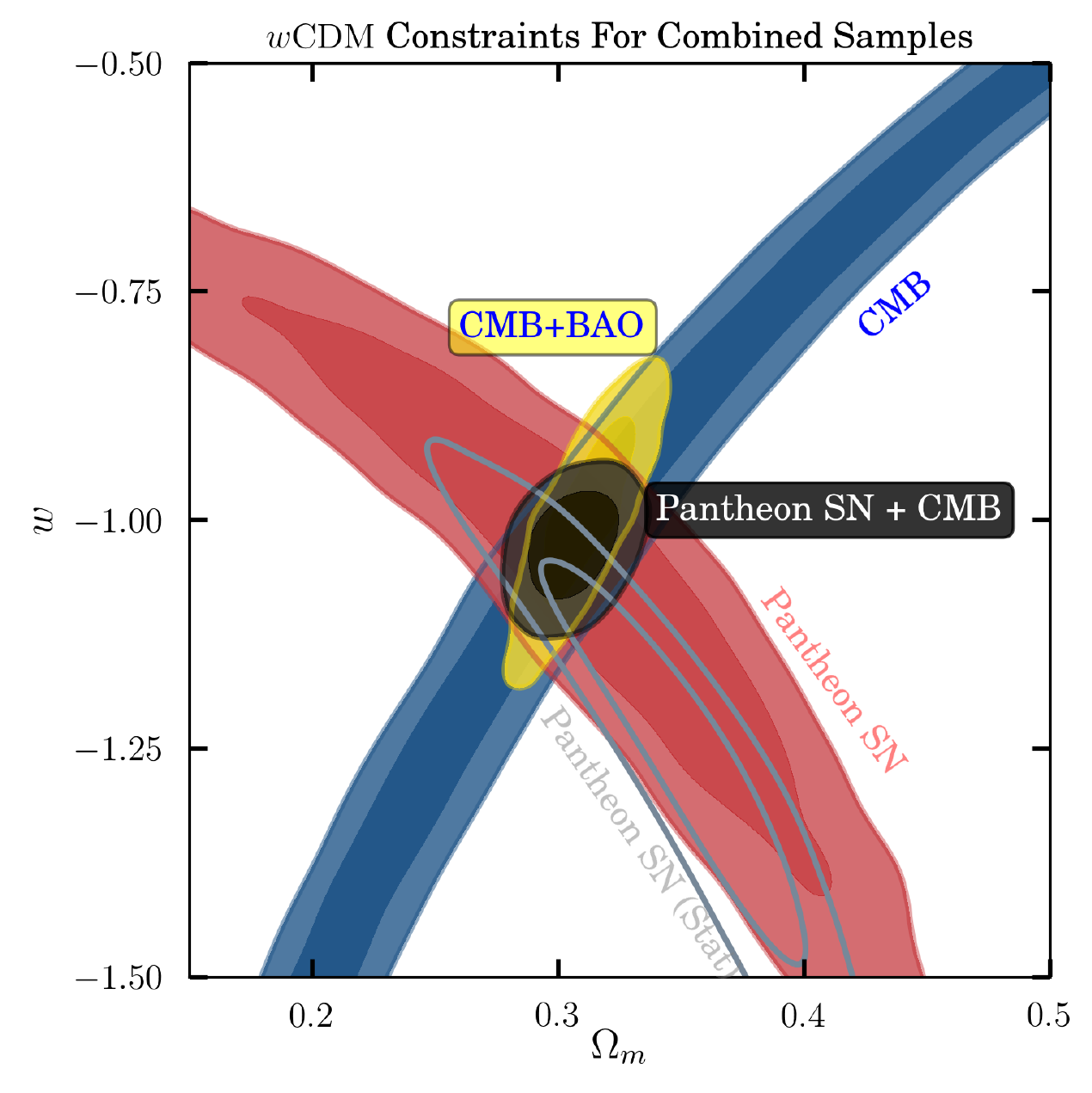}
\caption{Cosmological constraints derived from the Pantheon SNIa sample. Left: constraints on the $\ol-\om$ plane for a non-flat $\Lambda$CDM cosmology derived from the Pantheon sample in comparison with those coming from the R98 sample of \cite{riess_98}. Right: constraints derived from the Pantheon sample on the $\om-w$ plane for a CDM cosmology with a dark energy component with equation of state parameter $w$, in comparison with those coming from Planck CMB data, and with the combination of the two, and with BAO. Figure extracted from \cite{scolnic_18} ($\copyright$ AAS. Reproduced with permission).}
\label{fig:sn_constraints}
\end{figure}

The left panel of Figure~\ref{fig:sn_constraints} clearly shows that these new data have resulted in a significant improvement of the cosmological constraints, when compared with the original data of the High-z Supernova Research Team \cite{riess_98}. The Pantheon data \cite{scolnic_18} alone allow a high-significance detection of dark energy. When both statistical and systematic uncertainties are combined together, the result for a non-flat universe is $\ol = 0.733\pm 0.113$ ($6\sigma$ detection), while for $\ok=0$ the result is $\ol=0.702\pm 0.022$ ($32\sigma$ detection). This demonstrates the ability of SNIa observations to constrain cosmology on their own. However, much tighter cosmological constraints are achieved through the combination with other cosmological probes like the CMB or the BAO. The great advantage is that SNIa likelihoods are typically orthogonal to other measurements of cosmological parameters, the reason for this being the lower mean redshift coverage compared to most of other methods. This is crucial to break the geometric degeneracy of the CMB in the $\ol-\om$ plane (see in Figure~\ref{fig:planck_ok_om} this effect in the $\ok-\om$ plane). The combination of the CMB and SNIa also allows us to explore cosmological models with equation of state of dark energy $w\ne 1$ ($w=1$ corresponds to a cosmological constant), by breaking the degeneracy between $w$ and $\om$ (see right-panel of Figure~\ref{fig:sn_constraints}).

\section{Tensions, anomalies and open problems in the Standard Model}
\label{sec:tensions_anomalies}

The previous two sections have clearly shown that we nowadays have a very solid and well-established model that is able to describe the general properties of the Universe with just six parameters. It is quite remarkable that completely different and independent observations, spanning a wide range of redshifts (from the early to the present universe) agree to great precision on the values of these parameters. In the CMB itself, which is currently the probe giving the most stringent constraints on the model, there are three different observables (temperature anisotropies, polarisation anisotropies, and CMB lensing) giving fully consistent constraints. While the CMB temperature or polarisation anisotropies were not able to jointly fit the mass, curvature and dark-energy parameters (this is the well-known geometric degeneracy; see Figure~\ref{fig:planck_ok_om}), the CMB lensing, which has recently started to be exploited to constrain cosmological parameters, helps to alleviate this degeneracy. Combination with either BAO or SNIa observations totally breaks this degeneracy and, importantly, the result is the same independently on what of these two probes is used. 

This outstanding agreement between observations in the framework of a relatively simple 6-parameter model is a remarkable achievement, and has probably incited cosmologists to further scrutinise the data in the search for possible tensions or discrepancies that could point to new physics or yet undiscovered phenomena beyond the standard model. This search has led to some anomalies and tensions between some specific parameters, which lie at a not sufficiently high significance level as to be called ``discrepancies''. Apart from this, of course, it has to be born in mind that, while the model is well established, it points to the existence of two entities, dark matter and dark energy, which we currently do not understand. Neither we do properly understand the physics of inflation, which is a basic ingredient of the current model. These are open key problems that should drive the research in cosmology in the coming decades. In the following sections we will briefly describe these aspects.

\subsection{Tensions}
\label{sec:tensions}

\subparagraph{$H_0$ estimates in the local and in the early universe}

The tension that is currently being more hotly debated concerns the value of the Hubble constant, for which standard distance-estimation techniques in the local universe using SNIa (see section~\ref{sec:sn}) seem to give higher values than the one inferred from the parameters that best-fit the CMB anisotropies at $z=1100$ (see section~\ref{sec:cmb}). As it was commented before, the value derived by the SH0ES team from SNIa observations, $H_0=74.03\pm 1.42$~km~s$^{-1}$~Mpc$^{-1}$ \cite{riess_19}, differs at the $4.3\sigma$ level with the value derived from the Planck data, $H_0=67.66\pm 0.42$~km~s$^{-1}$~Mpc$^{-1}$ \cite{planck_18_6}. Independent local estimates rely on measuring time delays on strong-lensed QSOs. Applying this technique on six lensed systems the H0LiCOW team has recently reached a similar precision to other techniques on $H_0$, obtaining $H_0=73.3_{-1.8}^{+1.7}$~km~s$^{-1}$~Mpc$^{-1}$ \cite{wong_19}, a value that is in tension at the $3.1\sigma$ level with the CMB. When the two local estimates (time-delay cosmography \cite{wong_19} and distance ladder \cite{riess_19}) are combined, the difference with respect to the CMB increases to $5.3\sigma$. However, recently the CCHP team, calibrating SNIa distances using the TRGB method, came up with a lower value, $H_0=69.8\pm 1.9$~km~s$^{-1}$~Mpc$^{-1}$ \cite{freedman_19}, which alleviates the tension with the CMB.

The fact that different CMB experiments (Planck, ACT, SPT), on the one hand, and different local estimates (time-delays and SNIa), on the other hand, agree, is telling, and therefore the question of whether there is new physics beyond $\Lambda$CDM seems justified. Importantly, any proposed extension to the current model must alleviate the current tension but not at the cost of increasing the differences on other parameters, something that is not always easy to achieve. There are some recently proposed extensions to the model, like the existence of a scalar field acting as an early dark energy component \cite{poulin_19}, or a model with self-interacting neutrinos \cite{kreisch_19}. However, despite it is important to keep eyes open in the search for extensions to the current model that could lead to fresh discoveries, before extracting more firm conclusions it seems convenient that the data are examined in more detail in the search for possible systematics that could explain the differences. Also, future observations with higher precision will no doubt help to better establish the significance of the tension. More detailed discussions about this aspect can be found on the notes of a workshop recently held at the Kavli Institute for Theoretical Physics that was specifically dedicated to this topic \cite{verde_19}.

\subparagraph{$\sigma_8$ from galaxy cluster number counts and from the CMB}
Another apparent tension that has fostered interest on the last years concerns the joint $\om-\sigma_8$ estimates coming from cluster number counts and from the CMB, as illustrated in Figure~\ref{fig:om_s8_clusters} and previously referred to in section~\ref{sec:galaxy_clusters}. The tension initially pointed out in the Planck 2013 cosmological results (using DR1) is partially relieved in the Planck 2015 results (DR2) thanks to the lower value of $\sigma_8$ derived from the CMB, which in turn is due to the lower value of $\tau$. Using the CCCP value for the mass bias $(1-b)$, which is the default method in the Planck analysis \cite{planck_16_24}, the tension with the CMB is at $1.5\sigma$. This tension is not unique of Planck, but is also present in the cosmological analyses extracted from the number counts of other SZ surveys like ACT \cite{hasselfield_13} or SPT \cite{bocquet_15}, although typically at a lower significance because they are based on smaller cluster samples which leads to larger error bars. Currently the major uncertainty on these analyses is the mass bias estimate. According to the Planck 2015 analysis the value of the mass bias that is required to reconcile the CMB and number counts estimates is $(1-b) = 0.58\pm 0.04$ \cite{planck_16_24}, which means that clusters should have mass $\approx 40\%$ lower than derived from hydrostatic equilibrium estimates. While current numerical simulations and weak lensing estimates agree that the mass bias should instead be $(1-b)\approx 0.8$, there could be aspects related to incorrect modelling of cluster physics or systematics on the data that could explain this difference. 

If these possible systematics are overlooked, we might start considering extensions to the $\Lambda$CDM model that could explain the differences between the low- and high-reshifts determinations of $\sigma_8$. The most obvious one would be a non-minimal sum of neutrino masses. This scenario was considered in \cite{planck_16_24}, and the conclusion was that in fact while this led to a reduced tension on $\sigma_8$ it was at the cost of increasing the tension between other parameters. Another possibility discussed in \cite{planck_16_24} is that baryonic physics may influence the late-time evolution of density perturbations in such a way that feedback from active galactic nuclei could damp growth and therefore reduce $\sigma_8$. In any case, before developing any further these or other theoretical interpretations, it may be advisable to deepen our understanding of possible systematics associated with this analysis. As an example, recently \cite{remazeilles_19} pointed out that neglecting the relativistic corrections to the SZ frequency spectrum could lead to significant biases on the determination of $\sigma_8$ from the SZ power spectrum. And, more  importantly, a more precise determination of the mass bias parameter will give the definitive answer about the real tension between the CMB and cluster number counts. Improved cluster mass estimates from the CMB lensing, in the short term, or from future lensing surveys like Euclid, WFIRST and LSST that may allow us to reach $1\%$ precision in the determination of the mass bias parameter, will definitely contribute to this. 

\subparagraph{Planck low- and high-multipole data}
Although overall there is a quite remarkable consistency between the results of different CMB experiments, when they are analysed in closer detail some slight differences show up. Compared to WMAP results, Planck prefers a somewhat lower expansion rate, higher dark matter density and higher power spectrum amplitude, as it has been discussed in several Planck papers \cite{planck_16_13}, and also in \cite{addison_16}. It seems that these differences arise from the different multipole ranges that are sampled by Planck and WMAP. In fact \cite{addison_16} noted that there were some parameter shifts, in some cases at around $2-3\sigma$, between the $\Lambda$CDM parameters derived using Planck multipoles $\ell<1000$ or $\ell>1000$. They also pointed out some differences between high-$\ell$ Planck and SPT data, concluding that the  previous tensions may not be due to new physics but rather to systematics not accounted for in the data. Furthermore, none of the extensions that have been discussed in the literature seem to be preferred by the data. These issues were revisited in detail by \cite{planck_pip51_17}, and the conclusion was that, given the dimensionality of the model, the differences found between parameters are not statistically significant, and Planck low-$\ell$ and high-$\ell$ data are consistent with each other within around 10\% PTE.

\subparagraph{High amplitude of the lensing potential in the Planck TT power spectrum}
As it was commented in section~\ref{sec:cmb}, thanks to the increased quality of current datasets, the lensing of CMB photons by the LSS of the universe can now be used to constrain cosmological parameters. The lensing causes different effects in the CMB: a smoothing of the acoustic peaks and troughs in the temperature and polarisation power spectra, a conversion of E-modes into B-modes, and a generation of significant non-Gaussianity that can be measured through the CMB 4-point functions. Planck data show some differences in the amplitude of the lensing potential derived from the former and from the latter of these estimates, which are at around the $2\sigma$ level \cite{planck_16_13}. The difference in the combination of parameters $\sigma_8\om^{0.25}$, which is a good proxy for the lensing amplitude, is just $1.3\sigma$ \cite{planck_pip51_17}. However, as initially pointed out by \cite{addison_16}, this difference increases to $2.2\sigma$ \cite{planck_pip51_17} if only multipoles $\ell>1000$ are used to quantify the effect of the smoothing of the CMB peaks.

\subsection{Anomalies} 
Although there is an excellent consistency between the CMB data and the $\Lambda$CDM model, several features or anomalies in the CMB maps and power spectra have extensively been discussed in the literature, with the aim to precisely assess their significance and understand if there are really hints for extensions of the $\Lambda$CDM model. One of the CMB anomalies that has been more hotly debated is a lack of power and correlation on large angular scales, which shows up both in the power spectrum at multipoles $\lesssim 40$ (see Figure~\ref{fig:planck_ps}) and in the two-point angular correlation function. This lack of power is actually the main driver of the difference between the best-fit parameters from the low-$\ell$ and high-$\ell$ power spectra that were previously commented in section~\ref{sec:tensions}. Particularly low is the quadrupole amplitude, as first hinted at in COBE data. \cite{bennett_11} analysed this effect in WMAP-7yr data and found that its statistical significance was quite sensitive to the assumed foreground mask, to the statistical estimator, and also to the impact of the Integrated Sachs-Wolfe (ISW) on large angular scales. They also pointed out that the WMAP angular correlation function is consistent with the $\Lambda$CDM expectation within $2\sigma$ for all angular scales, and then downplayed the significance of this effect. Its significance is however larger (between $2.5\sigma$ and $3\sigma$, depending on the estimator used) in Planck data, due to the fact that Planck power spectrum is lower than the one from WMAP \cite{planck_13_15}. Remarkably, artificially scaling up the low-$\ell$ data to better match the best-fit model leads to an even lower value of $H_0$, which aggravates the tension with SNIa data (see section~\ref{sec:tensions}).

Another well-known anomaly is the ``cold spot'', a non-Gaussian negative feature seen in the CMB around $l=209^\circ$, $b=-57^\circ$ with a size of $\sim 5^\circ$, first identified in WMAP-1yr data using a spherical Mexican hat wavelet analysis \cite{vielva_04}, and with a significance of around $1.4-2.3\sigma$ \cite{bennett_11}. This feature was confirmed by Planck data, with a probability $\lesssim$1\%, the exact value depending on the exact size of the filter used \cite{planck_13_23}. Other two anomalies, also detected in WMAP-1yr data, and that have subsequently been extensively scrutinised are: a remarkable alignment between the directions of the quadrupole and the octopole, which are supposed to be randomly oriented; and a hemispherical asymmetry between the low-$\ell$ power spectra calculated in two halves of the sky separated with respect to the position $l=237^\circ$, $b=-20^\circ$, with significance between 95\% and 99\%. A more detailed overview of these an other anomalies can be found in different reviews \cite{schwarz_16,scott_18} or in the relevant WMAP \cite{bennett_11} and Planck \cite{planck_13_23} papers.

Currently there is no discussion that these anomalies do exist, both in WMAP and Planck data, this in fact being a good consistency cross-check of two different independent datasets, nor there is debate about their statistical significance. The question is whether these anomalies in the data may be connected or not with cosmic anomalies in the model. While some groups emphasise that we still lack an understanding of these large-scale features that seem to violate statistical isotropy and the scale invariance of inflationary perturbations \cite{schwarz_16}, there is general consensus that they do not provide sufficiently high statistical significance as to justify the endeavour of exploring extensions of the current $\Lambda$CDM model \cite{bennett_11}. The key issue has to do with the interpretation of the statistical significance of the different tests that are applied to the data. \cite{scott_18} points out that the statistical significance of all these features is usually assessed after they were actually discovered, disregarding other anomalies that could have been found and actually were not (this is what statisticians refer to as ``multiplicity of tests''). As an example \cite{scott_18} mentions that the statistical significance of the cold spot is usually assessed by looking to the probability of finding features with just the same angular size and amplitude, while we should instead marginalise over the scale as well as over the potential filter shapes. Of course, there is not a fully objective way to look for these effects. Therefore, in order to account for them, \cite{scott_18} recommends focusing on $5\sigma$ anomalies, and this would inevitably downplay the significance of the previous effects that is typically $2-3\sigma$. Also, it has to be taken into account the correlation between different anomalies when assessing their combined significance.

\subsection{Open problems}
In the previous sections we have described our current understanding of the Universe, which essentially is that it evolved from an early hot and dense phase, leading to the later hierarchical assembly of galaxies, clusters and superclusters at our epoch. However, the successes of the current $\Lambda$CDM model must inevitably be balanced by the fact that they rest upon three unknown or yet not well understood entities beyond the Standard Model of particle physics: inflation, dark matter and dark energy. These are the main mysteries in contemporary cosmology and, arguably, in all of physics.

The theory of inflation is indirectly supported by various observational pieces of evidence. It was initially proposed in the 80s to resolve a number of puzzles of standard big bang cosmology, such as the entropy, flatness, horizon, smoothness and monopole problems \cite{planck_13_22}. More importantly, inflation provides the seeds for structure formation. During inflation initial quantum fluctuations were stretched, thence leading to the cosmological fluctuations that later grew and, by gravitational instability, generated the structure of today's universe. Several predictions of inflation have been successfully confirmed by observations, namely that the Universe should have a flat geometry, nearly scale-invariant perturbations, and nearly Gaussian perturbations on all scales. However, at present we can not assert that any of the available observations prove that inflation is actually correct, and for this reason it remains being a theoretical framework rather than a model. We currently lack a precise understanding of how exactly inflation occurred, of the physics  of extremely high energies that are required to drive this period, nor we have a unique scenario of inflation. It is recognised that nowadays the best way to attain advances in this field is through the detection of the faint B-mode pattern in the CMB polarisation which should have been created by the gravitational waves generated during inflation (see section~\ref{sec:cmb} and Figure~\ref{fig:scalar_tensor_ps}). The detection of this signal would help constraining the energy scale of inflation -a key discriminant between different inflation models- and its dynamics, would in fact confirm the fourth prediction of inflation, which is the generation of a spectrum of tensor modes (gravitational waves), and would help measuring the exact shape of this spectrum. This is the main goal of the ongoing and planned CMB experiments.

As it was explained in the previous sections, there is compelling evidence that we have knowledge of just 5\% of the total mass-energy content of the universe, while 26\% is in the form of dark matter and 69\% in the form of dark energy. Current observations provide strong evidence for a non-baryonic nature of dark matter. Current evidence for the existence of non-baryonic matter come from oscillation experiments, which have shown that at least two of the three neutrino species have non-zero mass. However, CMB and LSS observations indicate that they only make up a small fraction of the dark matter in the Universe. In parallel to direct dark-matter searches in experiments like LHC, astrophysics and cosmology can provide insightful information about the nature of dark matter. Measuring the exact value of the total absolute neutrino mass, which can be done by measuring the tiny suppression of structure formation produced by neutrinos, is important to understand the mechanisms that gave them their mass. On the other hand, measuring the density profiles of dark matter halos and the matter power spectrum with very high precision can help to set constraints on the dark matter mass. This will be the goal of future LSS surveys

The origin of cosmic acceleration, which implies the existence of some distributed component of energy density not associated with matter concentrations and which exerts negative pressure, also remains a deep conundrum. One of the key issues to advance our understanding of the  nature of dark energy concerns the measurement of its equation of state, as it was commented in section~\ref{sec:dark_energy}. As it was described in section~\ref{sec:obs_probes}, all current measurements are consistent with a cosmological constant, i.e. vacuum energy with $w=-1$, and do not show any hint for a time-variation of this value. However, particle theories suggest that the dark energy density should not be constant and may have varied by many orders of magnitude over the history of the Universe. This would explain why its current value is so small, while the generally accepted theory of strong interactions (quantum chromodynamics) makes a contribution to the vacuum energy that is over 50 orders of magnitude larger than the observed value. Measuring the value of $w$ to higher precision, and constraining a possible time-variation of this parameter, is one of the main goals of future LSS surveys like Euclid, DESI or LSST. 

\section{Concluding remarks}
In this chapter we have shown that we currently have a consistent model describing the basic properties of the Universe we live in, its origin, evolution and matter-density content. This model is supported on Einstein's General Relativity, and states that the universe originated in a hot-dense phase called Big Bang, it underwent an early phase of expansion that created light elements via the big bang nucleosynthesis and the CMB, and a much earlier phase of accelerated expansion, called inflation, during which the initial quantum fluctuations were stretched to cosmic sizes, leaving an imprint on the CMB in the form of its anisotropies, and also leading to the formation of LSS due to gravitational instability. This model is strongly supported by a variety of observations, of which here we have described the most important ones: CMB, the LSS and Type Ia supernovae. Technological advances over the last decades have allowed us to exploit these cosmological probes to unprecedented precision, allowing cosmological parameters to be inferred with precision better than 1\%, and leading to what we have come to call the ``era of precision cosmology''. It is also an outstanding achievement that totally different and independent cosmological probes, like the CMB, the LSS or SNIa, give consistent constraints on the cosmological parameters. In fact, even using only the CMB, the data from the Planck satellite have allowed us to obtain consistent constraints from three different observables: temperature anisotropies, polarisation anisotropies, and the lensing of CMB photons.

Current observations are nicely described by a set of just 6 parameters, three of which describe the composition and evolution of the Universe, two the initial conditions of the density perturbations, while the sixth is an astrophysical parameter related to the optical depth of Thomson scattering to recombination. This model has been coined $\Lambda$CDM, to emphasise that it is dominated by a dark energy component, which causes the observed accelerated expansion that seems to be associated with a cosmological constant $\Lambda$ (i.e. vacuum energy with equation-of-state parameter $w=-1$) with a density parameter $\ol=0.69$, and by a cold dark matter component with $\oc=0.26$, while ordinary baryonic matter contributes to just $\ob=0.05$. The spectral index of the density perturbations is found to be $n_{\rm s}=0.9646\pm 0.0042$, which is a small deviation, at the $8\sigma$  level, from perfect scale invariance ($n_{\rm s}=1$), in agreement with the simplest inflationary models. Extensions from this model have been explored using existing data, but finding no compelling pieces of evidence. For instance, when different probes are combined no hints have been found of deviations from a perfectly flat geometry, with the best constraint $\ok = 0.0007\pm 0.0019$, coming from a combination of CMB and BAO, being astonishingly consistent with $\ok=0$. No evidence has been found either for a running of the spectral index accounting for a scale-dependency of the primordial fluctuations, as is predicted by some inflationary models. The equation of state of dark energy is found to be fully consistent with $w(a)=-1$, with no evidence of any time-variation of this value. The upper limit on the neutrino mass, as well as the number of relativistic species, are also fully compatible with the standard values. These results do not necessarily imply that these deviations do not exist, but rather that the existing data do not favour any of them within their error bars, and for this reason we can firmly assert that $\Lambda$CDM is the model that best describe these data.

Despite the overall stunning consistency between different and independent datasets, there are a few claimed tensions in some specific cosmological parameters. The most significant one, and currently more hotly  discussed, concerns the expansion rate. SNIa-based distance-estimation techniques in the local Universe give a value for $H_0$ that differs with the CMB measurement at $z=1100$ at $4.3\sigma$. If the SNIa measurements are combined with other local estimates based on time delays on strong-lensed QSOs, then the discrepancy with the CMB increases to $5.3\sigma$. Several ideas about extensions of $\Lambda$CDM that could restore concordance between the two measurements have been recently explored in the literature, but none of them seem to provide a convincing solution, either because in some cases a better agreement is achieved thanks to enlarging the uncertainties, or in some other cases at the cost of increasing the discrepancies in other parameters. Other important tension concerns the value of the amplitude of the scalar fluctuations, $\sigma_8$, derived from the CMB and from the number counts of galaxy clusters. In this case the discrepancy is at $\sim 1.5\sigma$, but the exact value is strongly dependent on the value assumed for the mass bias parameter. A more firm assessment of the real discrepancy requires a more reliable estimate of this bias parameter, for which currently different mass estimates give a considerable scatter. Therefore, it seems clear that before continuing to explore extensions to the current $\Lambda$CDM model, it may be better to wait until systematics affecting current analyses are reduced with the help of additional datasets with better sensitivity. An example are the new cluster mass estimates extracted from CMB lensing, that have just started to be exploited.

Finally, in spite of having a well consolidated model describing all current observations, we may not forget that two of its main ingredients, dark matter and dark energy, are currently unknown entities. Future LSS surveys will be key to shed new light on the nature of these entities, particularly important being the characterisation of the dark energy equation of state, and a possible time-variation of this parameter. This will be crucial to understand what is causing the observed accelerated expansion of the universe, or if alternatively our model requires modifications of General Relativity on cosmological scales. Equally important is understanding the generation of the primordial density fluctuations, for which the preferred model is inflation. Understanding the physics that drove this process could lead to important discoveries at energy scales not previously explored, and in this aspect ongoing and future CMB polarisation experiments will play a major role.

\begin{acknowledgement}
Most of of this work was written during a 5-week visit of the author to the University of Cambridge, in summer 2019. The author thanks the hospitality of the Cavendish Astrophysics group during this visit. The author also thanks John Beckman and Francisco-Shu Kitaura for reading parts of the text, and the referee for a careful reading of the text and useful comments. Some of the figures presented here have been taken from the ``Planck Image Gallery'' (ESA and Planck Collaboration).
\end{acknowledgement}

\bibliographystyle{spphys}
\bibliography{references.bib}

\end{document}